# Electron Transport in Dirac and Weyl Semimetals

Huichao Wang(王慧超)[1], Jian Wang(王健)[1,2,3,4]*

[1]International Center for Quantum Materials, School of Physics, Peking University, Beijing 100871, China

[2]Collaborative Innovation Center of Quantum Matter, Beijing 100871, China

[3]CAS Center for Excellence in Topological Quantum Computation, University of Chinese Academy of Sciences, Beijing 100190, China

[4]State Key Laboratory of Low-Dimensional Quantum Physics, Department of Physics, Tsinghua University, Beijing 100084, China

*Corresponding Author: jianwangphysics@pku.edu.cn

Recently, the Dirac and Weyl semimetals have attracted extensive attention in condensed matter physics due to both the fundamental interest and the potential application of a new generation of electronic devices. Here we review the exotic electrical transport phenomena in Dirac and Weyl semimetals. Section 1 is a brief introduction to the topological semimetals (TSMs). In Section 2 and Section 3, the intriguing transport phenomena in Dirac semimetals (DSMs) and Weyl semimetals (WSMs) are reviewed, respectively. The most widely studied $Cd_3As_2$ and the TaAs family are selected as representatives to show the typical properties of DSMs and WSMs, respectively. Beyond these systems, the advances in other TSM materials, such as $ZrTe_5$ and the $MoTe_2$ family, are also introduced. In Section 4, we provide perspectives on the study of TSMs especially on the magnetotransport investigations.

# 1. Introduction

In 2016, for theoretical discoveries of topological phase transitions and topological phases of matter, Thouless, Haldane and Kosterlitz were awarded the Nobel Prize in physics. By introducing the topological concepts from mathematics, they opened the door on a new world in physics where new and exotic phases of matter emerge. On the basis of their remarkable work, great progress has been made in the theoretical understanding and experimental realization of topological states over the past decade. In particular, topological materials have attracted widespread attentions since the discovery of topological insulators (TIs) [1-8] and become an important frontier in condensed matter physics.

Recently, the focus has moved towards topological semimetals (TSMs) since the theoretical proposal of a three-dimensional (3D) Weyl semimetal (WSM) in a magnetic phase of pyrochlore iridates [9-16]. TSMs are characterized by the symmetry protected band crossing in the Brillouin zone (BZ) at or close to the Fermi level. Based on the degeneracy and momentum space distribution of the nodal points, the currently most studied TSMs can be classified into three classes (Fig. 1): Dirac semimetal (DSM), WSM and nodal line semimetal (NLSM) [11]. The distinct electronic structure of TSMs gives rise to intriguing properties, which are important due to both the fundamental interest to explore new quantum phenomena and the potential application of a new generation of electronic devices.

In DSMs, the band crossings are isolated points with four-fold degeneracy, known as Dirac points (DPs) or nodes [11,15,16]. Graphene is an example of two-dimensional (2D) DSMs, in which the DPs are fragile. As the analogy of 3D graphene, 3D DSMs possess bulk DPs that are robust against perturbations [14]. The 3D DSMs are expected not only to show similar exciting phenomena and applications found in graphene but also to exhibit unusual properties such as giant diamagnetism and oscillating quantum spin Hall effect with the thickness [15-17]. In addition, the DSMs are at the boundary of various topological phases and can be regarded as the parent compounds to realize other topological states, such as WSMs and topological superconductors (TSCs) [17]. The DSM state has been sought by studying the topological phase transition in $Bi_{1-x}Sb_x$ [18], $TlBi(S_{1-x}Se_x)_2$ [19-21], $Hg_{1-x}Cd_xTe$ [22], $(Bi_{1-x}In_x)_2Se_3$ [23-25], $Pb_{1-x}Sn_xSe$ [26], $(Bi_{1-x}Sb_x)_2Se_3$ [24] etc., while the realization of a stable DSM phase in these systems is still challenging due to the requirement of fine-tuning composition. In 2014, the presence of stable DSMs was identified in $Na_3Bi$ and $Cd_3As_2$ by angle-resolved photoemission spectroscopy (ARPES) experiments [17,27-30] after the theoretical predictions from one group in China [15,16]. The experimental verifications ignite intensive studies in these two materials [31-36], where $Cd_3As_2$ has become the most studied DSM system due to its chemical stability in atmosphere. Besides, the DSM states were also suggested in other materials, such as $BiZnSiO_4$ family [37], BaAgBi family [38,39], $TlMo_3Te_3$ family [39], $PtBi_2$ [39], $Cu_3PdN$ [40] and the elemental α-Sn [41].

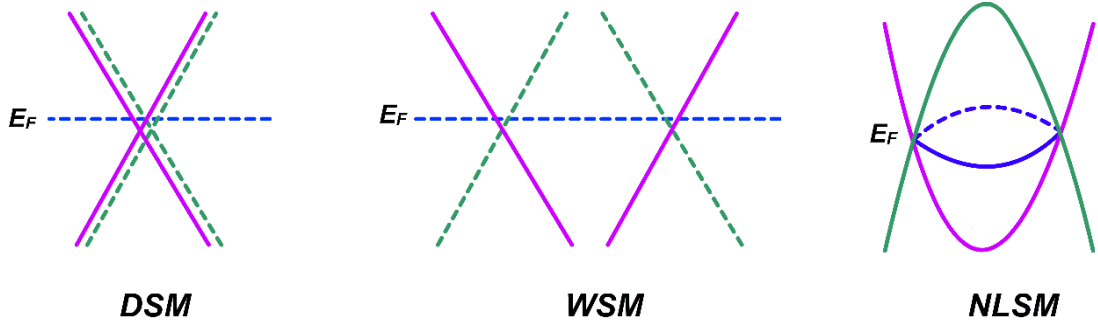

**DSM**  **WSM**  **NLSM**

Fig. 1. (color online) Schematic band structure (with Fermi level shift slightly off the nodal points) of the three kinds of topological semimetals: Dirac semimetal (DSM), Weyl semimetal (WSM), and Nodal-line Semimetal (NLSM). The two massless Weyl nodes with distinct chirality in DSM are slightly displaced for clarity.

For a WSM, the band crossings are isolated points with two-fold degeneracy, known as the Weyl points (WPs) or nodes, and play the role as monopoles in the Brillouin zone [9-11]. In fact, the Dirac node is composed of two Weyl nodes with opposite chirality overlapping each other in the momentum space (Fig. 1) [11]. WSMs feature the Weyl fermions as low-energy quasiparticles in the bulk and non-closed Fermi arcs (FAs) on the surfaces [9,10]. Weyl fermion is a massless chiral fermion, which has been long pursued as a fundamental particle in nature while the result is disappointing. The situation changes when considering the Weyl fermions as emergent quasiparticles in condensed matter systems [42]. In 2011, Wan *et al.* proposed the topological WSM state in pyrochlore iridates $R_2Ir_2O_7$ [9], and similar state was predicted to exist in the magnetically doped TI superlattice [10]. After that, a number of WSM candidates were then proposed [20,43-45] while the experimental confirmation of WSMs met difficulties until the discovery of such exotic states in TaAs family [46,47]. In 2015, two groups from China and USA almost simultaneously predicted that TaAs family materials are naturally WSMs [46,47]. The fascinating features of WSMs, Weyl fermion cones in the bulk and nontrivial FAs on the surfaces, were soon and independently confirmed in the TaAs family by the scientists from China, the United States and the United Kingdom by using ARPES technique [48-53], which stimulates intensive studies on the WSMs [54-68].

In a NLSM, two bands cross each other along a closed curve in the momentum space. The curve where the bands cross is called a nodal line, which may take the form of an extended line across the BZ, or a closed loop inside the BZ, or a chain consisting of several connected loops (nodal chain) [69]. In 2011, the NLSM was theoretically proposed to exist in a fine-tuned superlattice of a normal insulator and a TI with broken time-reversal symmetry (TRS) [70]. A series of realistic materials were then predicted to host the NLSM state, which include the Mackay-Terrones crystals [71], $PbTaSe_2$ [72,73], CaAgX (X=P, As) [74], $Ca_3P_2$ [75], $TlTaSe_2$ [76], $Cu_3PdN$ [40], etc. [11,69]. Recently, the experimental studies have presented some evidence for the NLSM phase in several bulk materials such as $PbTaSe_2$ [73], ZrSiX (X=S, Se, Te) [77-79], and $PtSn_4$ [80]. Moreover, the 2D nodal line fermions were also discovered in monolayer $Cu_2Si$ based on combined theoretical calculations and ARPES measurements [81].

In this chapter, exotic electrical transport phenomena in Dirac and Weyl semimetals under magnetic fields are mainly reviewed. In Section 1, a brief introduction to the history and the latest research progress of TSMs has been presented [82-89]. In Section 2 and Section 3, the magnetotransport investigations on DSMs and WSMs will be introduced, respectively. The observation of intriguing phenomena in these systems, such as the large magnetoresistance (MR), the ultrahigh mobility, the obvious Shubnikov-de Haas (SdH) oscillations, the chiral anomaly, etc. will be reviewed [33-36,59-61]. As the most widely studied materials, $Cd_3As_2$ and the TaAs family are selected as representatives to demonstrate the typical properties of DSMs and WSMs, respectively. Beyond these systems, the advances in other materials, such as the DSM $ZrTe_5$ and the type-II WSM $MoTe_2$ family, will be also introduced in the last part of the corresponding section (Section 2.6 and Section 3.4). In Section 4, we provide perspectives on the study of TSMs especially on the magnetotransport investigations.

## 2. Transport properties of Dirac semimetals

2.1 Ultrahigh mobility and giant magnetoresistance

In 2013, based on the first-principles calculations, Wang *et al.* proposed that $Cd_3As_2$ is a symmetry-protected DSM with a single pair of 3D Dirac points in the bulk and nontrivial Fermi arcs on the surfaces [16]. Subsequently, the direct experimental evidences for the confirmation of $Cd_3As_2$ as a DSM were obtained from the ARPES results [27-30], which immediately stimulates intensive studies on the properties of $Cd_3As_2$. Several groups from China and the United States, almost at the same time, soon reported the electrical transport properties of $Cd_3As_2$ bulk crystals, in which striking features like the ultrahigh mobility and the large MR are revealed [33-36].

In general, the single crystals of bulk $Cd_3As_2$ are synthesized from a Cd-rich melt in the evacuated quartz ampoule [90]. In the isolation from the finally resulting materials, two kinds of samples are mainly investigated. One is the needle-like single crystal with well-defined facets, in which the longest axis lies along [110] and the largest face is normal to [112]. The other is the large chunky multidomain crystals which lack defined facets. The crystal quality of $Cd_3As_2$ largely influences the transport properties, which might show a rich spectrum even when the samples are from the same batch [33-35]. For example, the mobility of $Cd_3As_2$ at 5 K can vary from $9\times10^6$ $cm^2V^{-1}s^{-1}$ to $4\times10^4$ $cm^2V^{-1}s^{-1}$ for different needle-shaped crystals, and the multidomain crystals show lower mobility of $1\sim2\times10^4$ $cm^2V^{-1}s^{-1}$.

In Liang *et al.*'s report, the mobility of the DSM $Cd_3As_2$ can be as high as $9\times10^6$ $cm^2V^{-1}s^{-1}$ at 5 K for certain needle-shaped crystal [33]. High residual conductivity was observed in zero magnetic field and can be rapidly removed by an applied magnetic field. Based on the observation, the ultrahigh mobility was suggested to arise from a remarkable mechanism, which strongly protects the carriers against backscattering at zero magnetic field and results in a transport lifetime $10^4$ times longer than the quantum lifetime [33]. However, the physical nature and details of the protection mechanism are not very clear. The ultrahigh mobility of high quality $Cd_3As_2$ single crystals was also demonstrated in Zhao *et al.*'s report [35]. In one of the needle-shaped $Cd_3As_2$ samples, a conservatively estimation on the mobility gives the value to be $9.19\times10^6$ $cm^2V^{-1}s^{-1}$ and the mean-free path is about 0.25 mm at 6 K, which are consistent with Liang *et al.*'s results. The high mobility and long mean-free path of the DSM may bring various applications on future functional devices.

In a magnetic field, a very large MR $\rho_{xx}(15T)/\rho(0) = 1,600$ at 2.5 K (H ∥ [112]) can be observed in certain needle-shaped $Cd_3As_2$ crystals [33]. Liang *et al.* interpreted the giant MR as a result of the break of the protection mechanism that strongly suppresses backscattering and may be related to the changes of the Fermi surface under an applied magnetic field [33]. Besides, a pronounced linear MR was observed in the relative low-mobility samples [33]. He *et al.* reported similar linear MR behavior in $Cd_3As_2$ single crystals and showed that the large linear MR could survive even at near room temperature [34]. Feng *et al.* synthesized single crystals of $Cd_3As_2$ by using the chemical vapor transport technique and investigated the linear MR in details in the Hall-bar device of $Cd_3As_2$ thin plate with very low carrier densities [36]. They reported a large linear MR with $\rho_{xx}(15T)/\rho(0) = 31$ at 2 K and suggested that the field-induced relative shifting of the two Weyl-Fermi surfaces in each Dirac cone in momentum space is likely responsible for the linear MR behavior.

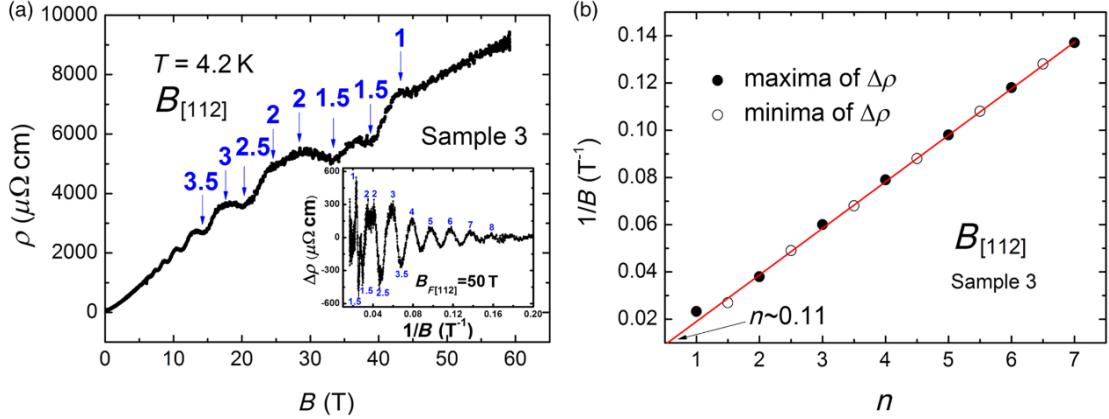

Fig. 2. (color online) (a) Magnetoresistivity of $Cd_3As_2$ up to 60 T. Inset: SdH oscillations after subtracting a polynomial background. The linear MR beyond the quantum limit (n=1) is a reminiscence of the Abrikosov's quantum MR model. (b) The Landau fan diagram of SdH oscillations. [35]

In the light of the Abrikosov's theory [91], a quantum linear MR can be expected to occur beyond the quantum limit (QL) in a gapless semiconductor with linear energy dispersion, where all the carriers occupy the lowest Landau level. Though the linear MR behaviors have been observed in $Cd_3As_2$ as mentioned above, their origins cannot be attributed to this scenario since the observed linear MR exists at very small magnetic fields [33,34,36]. Almost at the same time, Zhao *et al.* studied the MR of $Cd_3As_2$ at an ultrahigh magnetic field up to 60 T. With increasing magnetic field, the $Cd_3As_2$ single crystal exhibits obvious SdH oscillations and the Zeeman splitting appears (n = 2, 1.5) when the Fermi level is nearly pushed into the lowest Landau level. The sample reaches the QL at about 43 T. With further increasing magnetic field beyond the QL (B > 43 T), the $Cd_3As_2$ crystal presents an outstanding linear MR (Fig. 2(a)), which is reminiscent of the Abrikosov's quantum MR model.

In addition, nonsaturating linear MR persisting up to 65 T was also observed in the n-doped $Cd_3As_2$ system [92]. The MR variation was consistent with the results of the classical disordered model originating from fluctuating mobility, and thus it was suggested to be caused by the inhomogeneous mobility distribution [92].

2.2 Nontrivial Berry phase and anisotropic Fermi surface

On the giant MR of DSM $Cd_3As_2$ crystals, there is usually a superimposition of the SdH oscillations, which are the manifestation of the intrinsic Landau level properties. Analyses on the SdH quantum oscillations can obtain important parameters of the material, such as the mobility, the carrier density, the Dingle temperature, the transport time, etc. Besides, it is believed that the nonzero Berry phase, a distinguished feature of Dirac fermions, can also be identified by plotting the Landau fan diagram of the SdH oscillations. The oscillations can be described by the Lifshitz-Kosevich formula $\cos[2\pi(F/B+\gamma-\delta)]$, where $F$ is the oscillating frequency, $\gamma$ is the phase related to the Berry phase $\phi_B$ with $\gamma=1/2-\phi_B/2\pi$ and $\delta$ is a phase shift determined by the dimensionality [93]. A nontrivial $\pi$ Berry phase induces a zero $\gamma$ and the phase shift $\delta$ changes from 0 for a quasi-2D cylindrical Fermi surface to $\pm 1/8$ (+ for holes and − for electrons) for a corrugated 3D Fermi surface. Thus, the phase shift $\gamma-\delta$ can provide valuable information about the topography of materials.

For the $Cd_3As_2$ single crystals, He *et al.* observed a phase shift $\gamma-\delta \approx 0+0.06\sim0.08$ by analyzing the SdH oscillations at low temperatures [34]. The result was attributed to the nontrivial $\pi$ Berry phase, which provides evidence for the existence of Dirac fermions in $Cd_3As_2$. The slight deviation of 0.06~0.08 was suggested to from the correction of the 3D Fermi surface. The phase shift in the SdH quantum oscillations of $Cd_3As_2$ single crystals was also studied by Zhao *et al.* in details [35]. It was found that the intercept value may show a little shift under different magnetic field regimes, which suggests that the high magnetic field is needed to fix the intercept more reliable [35]. Zhao *et al.* measured the MR in an ultrahigh magnetic field up to 60 T and the Landau fan diagram of SdH oscillations gives an intercept of 0.11 (Fig. 2(b)) [35]. The intercept value deviates from the theoretical value of −1/8 for the n-type $Cd_3As_2$, which was attributed to the combined correction from the two Dirac points when the Fermi energy is above the Lifshitz saddle point. In this case, different intercept values might be obtained for different magnetic field orientations [35]. The positive phase shifts of electron carriers can also be explained by taking into account the inter-Landau band [94].

According to the numerical results, Wang *et al.* clarified that the resistivity peaks appear near Landau band edges and correspond to integer Landau indices. Moreover, the combined phase shift takes the discrete values of either ±1/8 or ±5/8 for DSMs from around the Lifshitz point to higher Fermi energies [94].

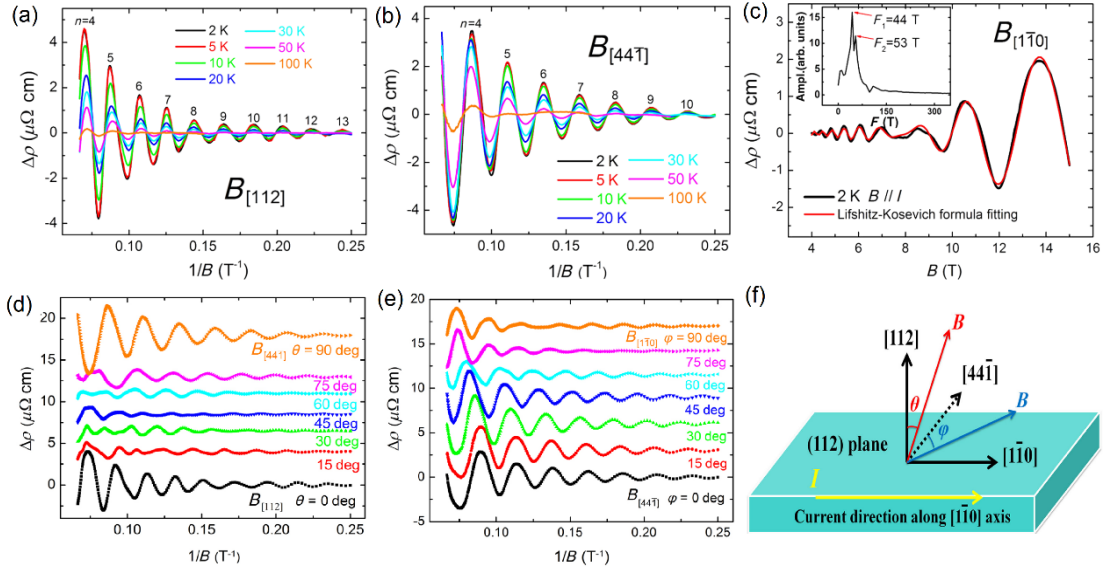

Fig. 3. (color online) (a-c) SdH oscillations in $Cd_3As_2$ single crystal at various temperatures when the magnetic field is applied along the [112], [44$\bar{1}$] and [1$\bar{1}$0] axis, respectively. (d,e) Angular dependence of SdH oscillations in $Cd_3As_2$ at different magnetic field directions. (f) Schematic structure for the magnetotransport measurements in $Cd_3As_2$. [35]

In a magnetic field, the SdH oscillations can be a very useful tool to detect the information of the Fermi surface. However, the magnetotransport measurements of the SdH oscillations in one direction or by rotating the magnetic field in a certain plane can only get a glimpse of a small part of the complete 3D Fermi surface of $Cd_3As_2$. An overall study on the transport properties of $Cd_3As_2$ at different magnetic field directions would be of interest for a comprehensive understanding of the DSM $Cd_3As_2$. Zhao *et al.* mapped an overview of the Fermi surface of $Cd_3As_2$ single crystals for the first time by systematic electrical magnetotransport measurements [35], complementary to previous ARPES experiments [27-30]. In the transport results, it was found that when the magnetic field lies in the [112] or [44$\bar{1}$] axis, MR oscillates with only one single frequency; however, the oscillations show two oscillating frequencies when the field is applied along the [1$\bar{1}$0] direction (Figs. 3(a)-3(c)). Further MR measurements by applying the magnetic field in different planes reveal that the two frequencies are also present when aligning the magnetic field at certain directions between the three axises (Fig. 3(d), Fig. 3(e), Table 1). The results of the SdH oscillations at varying crystallographic orientations were attributed to the sophisticated geometry of Fermi surface with two nested anisotropic ellipsoids around the DPs. In the studied $Cd_3As_2$ samples, the Fermi level locates slightly above the Lifshitz point, and thus the Fermi surface is formed by two nested ellipsoids which are evolved from the two separate anisotropic ellipsoids around the Dirac point. For a magnetic field with fixed direction, the SdH oscillations exhibit one single frequency when the maximum cross section of the Fermi surface does not pass through the nested region, and two oscillating frequencies will be observed when the nested region is involved. The geometry of the maximum cross sections in the Fermi surface can qualitatively interpret the frequency changes for the SdH oscillations in Table 1.

| Rotation angle | 0 deg | 15 deg | 30 deg | 45 deg | 60 deg | 75 deg | 90 deg |
|---|---|---|---|---|---|---|---|
| $B_{[112]}$ rotate to $B_{[44\bar{1}]}$ | 51 T | 44 T / 55 T | 51 T / 62 T | 51 T | 46 T | 43 T | 41 T |
| $B_{[44\bar{1}]}$ rotate to $B_{[1\bar{1}0]}$ | 41 T | 41 T | 44 T | 44 T | 46 T / 51 T | 45 T / 52 T | 44 T / 53 T |

Table 1 Fast Fourier Transform results of the SdH oscillations in DSM $Cd_3As_2$ single crystals under various magnetic field directions [35].

2.3 Low-dimensional Dirac semimetals

In addition to detect the transport properties of the DSM of bulk $Cd_3As_2$, it is also of interest to explore the intriguing physics in low-dimensional DSMs. For example, an oscillating quantum spin Hall effect is expected in the quantum well structure of DSMs, which means that the system shows oscillatory crossover between trivial and nontrivial 2D insulators as a function of the thickness. In addition, the investigations on the low-dimensional DSMs are crucial to the potential nanoelectronic applications. The synthesis of low-dimensional DSM $Cd_3As_2$ has been realized, in which $Cd_3As_2$ thin films were grown in a molecular beam epitaxy system by directly evaporating the bulk materials onto substrates [95,96] and high quality $Cd_3As_2$ microbelts, nanobelts, nanowires, etc. were fabricated by chemical vapor deposition (CVD) method [97-103].

The transport properties of the $Cd_3As_2$ thin films, tuned by an applied gate voltage, thickness or doping, were investigated. In a 50 nm-thick $Cd_3As_2$ film, Liu *et al.*

demonstrated a transition from band conduction to hopping conduction via electrostatic doping by solid electrolyte gating [95]. The extreme charge doping also enables an observation of a transition from electron- to hole-dominated two-carrier transport [103]. Besides, Cheng *et al.* revealed a thickness-induced semimetal-to-semiconductor transition in the single-crystalline $Cd_3As_2$ films [104]. In contrast with the metallic behavior in the bulk counterpart, the 50 nm-thick $Cd_3As_2$ film exhibits semiconducting characteristics, suggesting a bandgap opening. With the reduced film thickness from 900 nm to 50 nm, a transition from nontrivial to trivial Berry phase was also revealed. Furthermore, Yuan *et al.* investigated the quasiparticle dynamics and band evolution in $Cd_3As_2$ thin films by controlled Chromium (Cr) doping [105]. In undoped $Cd_3As_2$, the observed square-root-B relation of inter-Landau-level resonance signals the existence of massless Dirac fermions. With the controlled Cr doping, a topological phase transition and Dirac mass acquisition can be achieved, which enables a dynamic control of the quasiparticles.

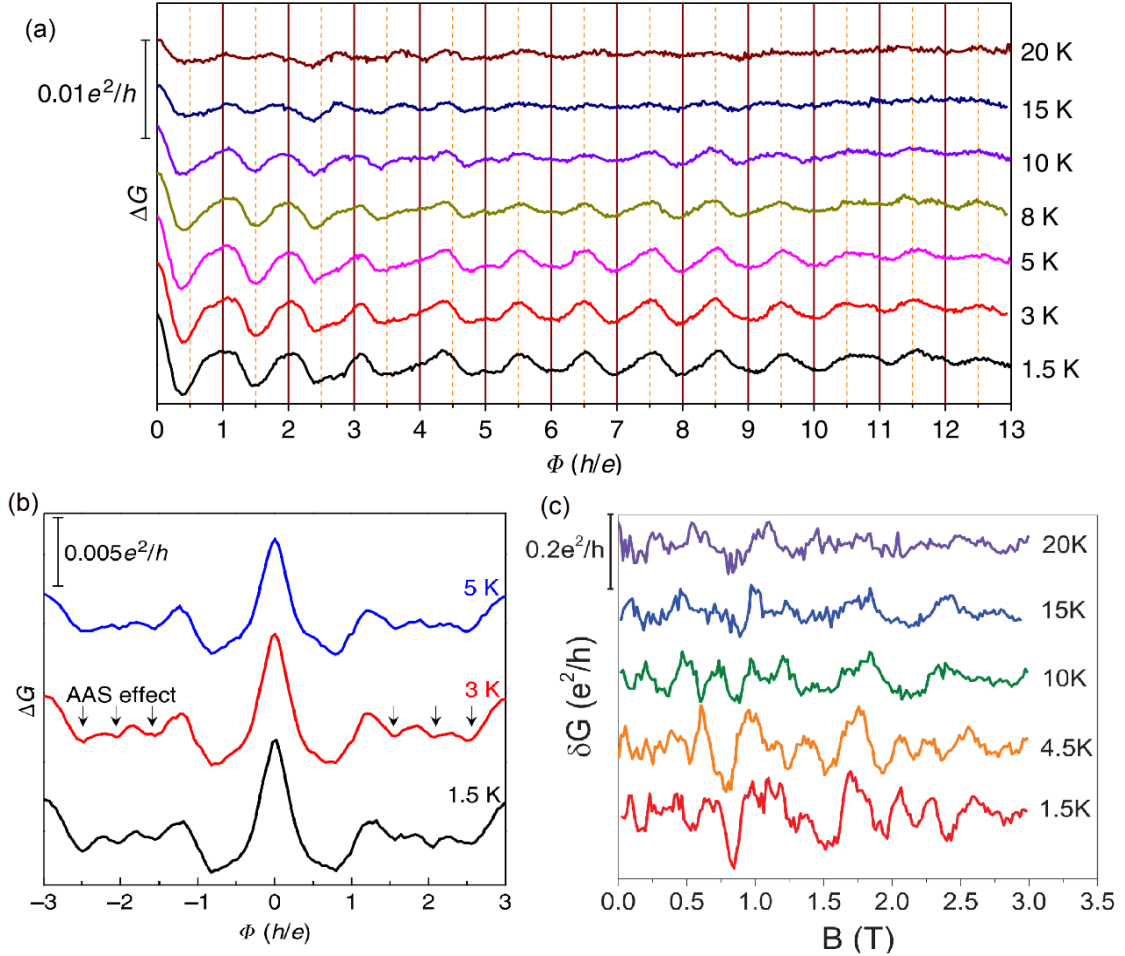

Fig. 4. (color online) (a) Aharonov–Bohm oscillations, (b) Altshuler–Aronov–Spivak oscillations and (c) universal conductance fluctuations in $Cd_3As_2$ nanowires. [106,107]

In the low-dimensional Cd$_3$As$_2$ structures, the high quality Cd$_3$As$_2$ microbelts and nanobelts grown by CVD exhibit the metallic behavior similar to the bulk crystal with very high mobility and pronounced SdH oscillations [98,99]. Besides, the semiconducting-like behavior was also observed in the resistance-temperature (RT) characteristic of some mesoscopic Cd$_3$As$_2$ structures, e.g. the thin films, nanowires and microribbons [95,99-101,108]. The semiconducting-like behavior might origin from the bandgap opening due to quantum confinement effect [99] or the low carrier density as the Fermi level is very close to the Dirac point [100]. Particularly, the magnetotransport measurements on the Cd$_3$As$_2$ nanowires reveal glaring quantum interference effects when the samples are scaled down to the phase-coherent region and thus the quantum correction to classical conductance becomes remarkable. Wang *et al.* reported the Aharonov–Bohm (AB) oscillations, Altshuler–Aronov–Spivak (AAS) oscillations and universal conductance fluctuations (UCF) in individual Cd$_3$As$_2$ nanowires (Fig. 4) [106,107]. In the nanowire ~115 nm in diameter, as increasing the magnetic field parallel to the nanowire direction, the conductance oscillation peaks at odd integers of h/2e with a period of h/e (Fig. 4(a)) signifying the AB effect. Besides, a phase shift in AB oscillations was observed and ascribed to the splitting of the DPs due to the magnetic field-induced TRS breaking. In the thick nanowire with a diameter of ~200 nm, oscillations with h/2e flux period were observed and ascribed to the AAS oscillations (Fig. 4(b)). Besides, the UCF was reported in the Cd$_3$As$_2$ nanowires with the diameter of 100 nm and 130 nm (Fig. 4(c)) when the magnetic field is perpendicular to the nanowire direction [107].

2.4 Topological phase transition to Weyl semimetals

The DSM is on the boundary of various topological phases [15-17]. It means that by modulation, the DSMs can be driven into other topological states, such as TIs, TSCs, WSMs, etc. Theoretically, the presence of a magnetic field would break the TRS and the overlapped WPs in the DSMs may be split into two separated WPs, which indicates a transition from DSMs to WSMs possessing chiral Weyl fermions and FAs. The experimental observation of a large spin-splitting of the conduction band in a magnetic field suggests a removal of spin degeneracy in DSM Cd$_3$As$_2$ by breaking TRS. Combining a band structure model consistent with the results of spin-splitting and extended Dirac-like dispersion, Jeon *et al.* reported that Weyl fermions might be the low-energy excitations in Cd$_3$As$_2$ when an external magnetic field is applied along the axis of the Dirac points [32]. Cao *et al.* also demonstrated the possibility of inducing WSM phase in DSM $Cd_3As_2$ at high magnetic fields by detecting the Landau level splitting and an angular dependent Berry phase [109]. An unexpected nontrivial Berry phase was obtained in the DSM $Cd_3As_2$ in certain measurement configuration when a field-generated mass term could be introduced to the system. Based on the observation, it was suggested that the results might serve as the hints for the emerging Weyl fermions in $Cd_3As_2$, where the spin degeneracy is removed by breaking TRS and a WSM phase is further demonstrated.

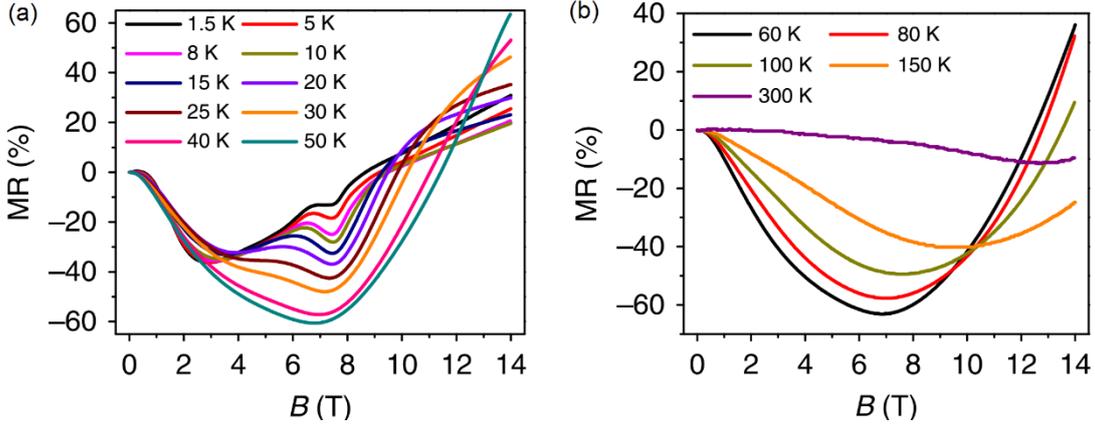

Fig. 5. (color online) (a,b) Negative MR in a $Cd_3As_2$ nanowire measured at temperatures from 1.5 K to 300 K [100]. The negative MR was ascribed to the chiral anomaly.

Chiral anomaly is a phenomenon associated with the chiral Weyl fermions (see more in Section 3.1) [110-115]. Since the Dirac node in DSMs is virtually two Weyl nodes with opposite chirality which can be split in the momentum space under an external magnetic field, the chiral anomaly is also expected to be observed in DSMs. The chiral anomaly in DSMs was firstly revealed in $Bi_{1-x}Sb_x$ (x ≈ 3%) near the critical point of the topological phase transition by the observed negative MR (NMR) phenomenon [98]. NMR evidence for the chiral anomaly in DSM $Na_3Bi$ was also presented [18,31]. Though the clear evidence of chiral anomaly has not yet been unveiled in bulk $Cd_3As_2$ since the NMR term in bulk $Cd_3As_2$ may be too small to be observed in the large MR background [33], the anomaly has been observed in various $Cd_3As_2$ nanostructures [100-102]. Li *et al.* reported a large NMR with magnitude of -63% at 60 K and -11% at 300 K in the single crystal $Cd_3As_2$ nanowires when the magnetic and electric fields are parallel, signaling the appearance of the chiral anomaly (Fig. 5) [100]. Besides, Li *et al.* detected an obvious

NMR in $Cd_3As_2$ microribbons in parallel magnetic fields [101]. The NMR show features (1) sensitive to the angle between magnetic and electrical fields, (2) robust against temperature and (3) dependent on the carrier density, which were attributed to the demonstrations of the chiral anomaly [101]. In addition, Zhang *et al.* reported the existence of the chiral anomaly in $Cd_3As_2$ nanoplates by presenting three independent experimental evidences, which includes results of the NMR, the valley transport, and the E·B-generated magneto-optical Kerr effect [102]. The work indicates that the chirality-polarized states are directly coupled with spin, orbit, and valley degree of freedom, which may stimulate investigations on the application in electronics.

Besides the appealing chiral anomaly associated with the Weyl fermions, the WSM has disconnected FAs on the surfaces due to the broken translational symmetry. In the presence of a static magnetic field perpendicular to the two opposite surfaces of a WSM, the disjointed FAs from two opposite surfaces can intertwine with chiral bulk modes and develop unusual closed magnetic orbits. A new type of quantum oscillation that arises from such Weyl orbits can in principle be detected [116]. For the DSMs consisting of two superposed copies of WPs, similar quantum oscillations might also occur while there are additional complications that must be considered, e.g. the reduced symmetry of the surfaces, the reduced symmetry in a magnetic field, and their interplay with bulk topology [116]. Experimental evidence for the Weyl orbit was claimed in focused-ion-beam-prepared microstructures of $Cd_3As_2$, in which an additional 2D quantum oscillation emerges when the thickness of the sample is comparable to the bulk transport mean free path [117]. However, Kargarian *et al.* argued that the double FAs in DSM is in fact unstable [118]. In contrast to the FAs in WSM, the doubled FAs in DSMs are not topologically protected in general and can be continuously deformed into a Fermi pocket in the presence of a surface perturbation. Thus, the reported Weyl orbits in DSM $Cd_3As_2$ microstructures might be a consequence of the mixing of bulk and surface states [118]. Recently, Zheng *et al.* suggested an observation of the quantum oscillations associated with such Weyl orbits in $Cd_3As_2$ nanoplates [119]. When the thickness of the nanoplates is smaller than the quantum mean free path of the electrons, additional 2D MR oscillations were observed at high magnetic fields to be superimposed on the conventional 3D bulk oscillations. The oscillating frequency of the 2D oscillations is close to the theoretical calculation of the Fermi-arc oscillation frequency. Thus, the discovery was ascribed to the manifestation of the Weyl magnetic orbits connecting the open FAs from opposite surfaces. Moreover, further evidence detected by the nonlocal transport measurements was also presented in the work to unveil the transport properties of FAs in TSMs. The 2D surface transport in the DSM $Cd_3As_2$ was also studied by Zhang *et al.* based on a series of $Cd_3As_2$ nanoplates [120]. In the selected different samples with lowering Fermi level, it was found that the contribution from bulk oscillations decreases and the surface state transport dominates gradually at the perpendicular magnetic fields. With further increasing magnetic fields, the system was driven into an unconventional quantum Hall state [120], which was also observed in thin films of the DSM $Cd_3As_2$ [121,122]. The origin of the quantum Hall effect and its relation to the Weyl orbits are calling for further investigations.

2.5 Superconductivity in Dirac semimetals

The DSMs can be driven to many other phases by breaking certain symmetries, in which the topological phase transition from a DSM to a WSM has been introduced [32,105]. Besides, a very attracting topic arises from the possible transition from a DSM to a TSC [1,2,123,124]. TSCs show a superconducting gap in the bulk state, but support gapless Majorana fermions or Majorana zero modes in the boundary [125-128]. The Majorana zero modes obey non-abelian statistics, and can be applied to topological quantum computation and to build a fault-tolerant quantum computer [129,130].

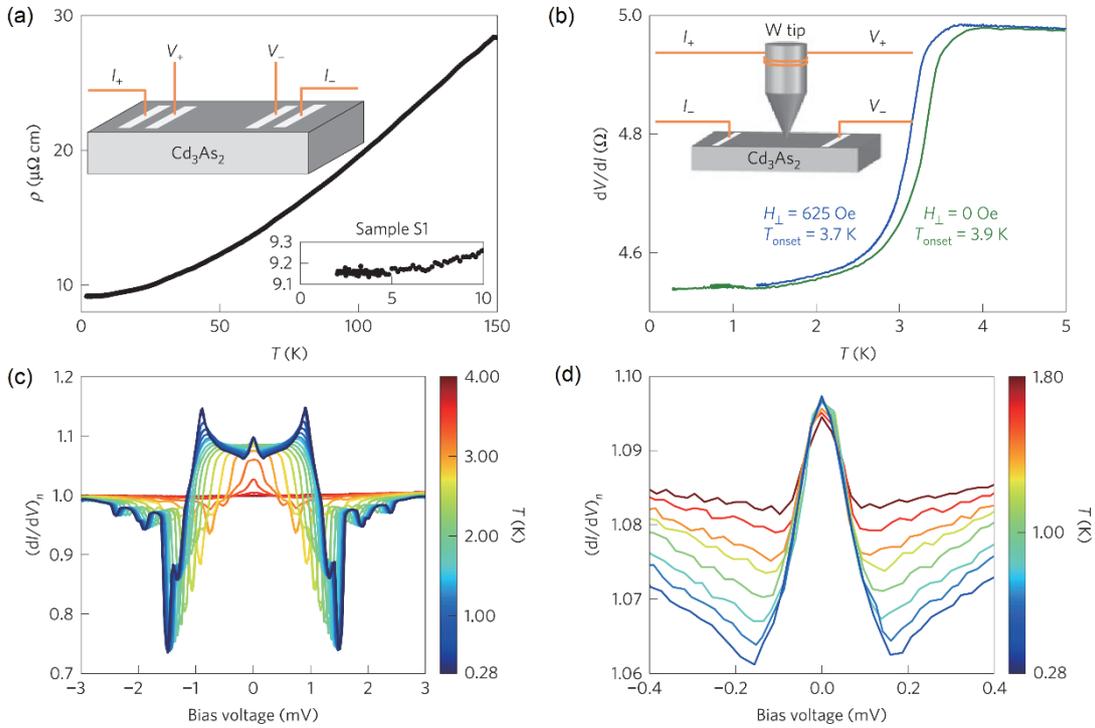

Fig. 6. (color online) (a) RT characteristic of the bulk $Cd_3As_2$. Upper Inset: Schematics of the standard four-probe method measurement configuration. Lower inset: Zoom-in of the RT curve below 10 K. (b) Temperature dependence of the zero-bias PC resistance.

Inset: Schematics of the PCS measurement configuration. (c) Normalized dI/dV spectra at different temperatures. (d) Zoom-in of the normalized ZBCP. [131]

In 2016, Wang *et al.* discovered superconductivity induced by hard point contact (PC) on DSM $Cd_3As_2$ crystals (Fig. 6) [131]. A significant resistance drop was observed in the RT behavior in the hard PC (W tip) measurements on the intrinsically metallic $Cd_3As_2$ (Figs. 6(a) and 6(b)). The resistance drop was deemed as a signature of superconductivity since an applied perpendicular magnetic field suppresses the critical temperature. 'Soft' PC (silver paste attached Au wire) measurement was also performed in the work but the superconducting behavior was absent. Thus, the observed superconductivity was suggested to be induced by the "uniaxial" pressure or doping of the hard tip contact. In the measured PC spectra (PCS), the main features of dI/dV spectra can be classified into two types (Figs. 6(c) and 6(d)): (i) double conductance peaks (DCPs) and double conductance dips both symmetric to zero bias, with the peaks smearing into a broad hump at high temperatures; (ii) a zero-bias conductance peak (ZBCP) observed at low temperatures. Besides, the DCPs and double conductance dips were found to exist independently of the ZBCP because the ZBCP could survive at certain temperatures (e.g. 2 K) where the DCPs have disappeared. The exotic features were suggested to originate from topological superconductivity in the surface FAS and bulk states of DSM $Cd_3As_2$ and the ZBCP might arise from Majorana fermions or bound states, which were supported by the theoretical analyses. Considering the topological property of the DSM, the findings indicate that the $Cd_3As_2$ under certain conditions could be a TSC [132].

The PC measurement was previously used to study superconducting materials and get the information of superconductivity [133,134]. In Wang *et al*'s work, the hard PCS technique has been developed to induce superconductivity on non-superconducting topological materials by using non-superconducting tip (hard PC), different from previous PC experiments on superconductors. More importantly, the results reveal a new way to detect and study TSCs, which is different with the prevailing proximity effect method on superconductor-TI (spin-orbital coupling semiconductor) hybrid structures for creating topological superconductivity or Majorana fermions.

The exotic discovery of the superconductivity in DSM $Cd_3As_2$ promptly attracted attentions from both theorists and experimentalists. The suggestion that the pressure and doping induced by the hard tip may be the origin of the superconductivity in $Cd_3As_2$ was further confirmed by a later experimental work. He *et al.* presented the resistance measurements on $Cd_3As_2$ single crystals under pressure up to 50.9 GPa [135]. With the increasing pressure, the low temperature resistance of $Cd_3As_2$ firstly increases while finally decreases when the pressure is above 6.4 GPa. Superconductivity appears at 8.5 GPa with a $T_c$ of about 2.0 K. The $T_c$ keeps increasing to about 4.0 K at 21.3 GPa, then shows nearly as a constant up to the highest pressure 50.9 GPa. In addition, it is noted that another group also studied the polycrystalline $Cd_3As_2$ by hard PC measurements and observed the indication of superconductivity [136].

2.6 Beyond the $Cd_3As_2$ system

Besides the $Cd_3As_2$ system, the properties of DSM were also explored in many other materials, such as $Na_3Bi$ [31,137,138], $PtBi_2$ [139], and $Zr(Hf)Te_5$ [140-163]. The theoretical prediction of $Na_3Bi$ as a DSM was presented in 2012 [15], and soon the experimental results from ARPES confirmed the 3D DSM state in $Na_3Bi$ [17]. The $Na_3Bi$ material is incredibly unstable upon exposure to air and moisture, which hinds the wide investigations on this system. Nevertheless, using special care to prepare the samples, Xiong *et al.* revealed the chiral anomaly [31] and the anomalous conductivity tensor [137] in the DSM $Na_3Bi$. In addition, Hellerstedt *et al.* grew large-area thin films of $Na_3Bi$ and investigated the electronic properties by back gating method [138].

The hunt for new natural DSM materials with novel properties has been an important issue since the successful discovery of DSM $Cd_3As_2$ and $Na_3Bi$. Gibson *et al.* predicted a series of compounds as the DSM candidates [39], which includes the Pyrite-type $PtBi_2$. Gao *et al.* synthesized high-quality $PtBi_2$ single crystals and carried out detailed magnetotransport measurements [139]. They observed an extremely large unsaturated MR up to $11.2 \times 10^6$ % at 1.8 K in a magnetic field of 33 T, which is higher than any other known Dirac materials reported before. Besides, significant SdH quantum oscillations at low temperatures suggest a nontrivial Berry phase in the $PtBi_2$, and the Hall analysis reveals ultrahigh mobility of the material. The experimental observations are consistent with the ab initio calculation in the work, which provides evidence that $PtBi_2$ can be a topological semimetal with Dirac point around the Fermi level.

The DSM state was also studied in topological nontrivial $Zr(Hf)Te_5$ [140-166], a layered material with van der Waals coupling between individual layers. The compound was initially studied for its resistivity anomaly, thermoelectric properties and quantum oscillations [167-173]. The second study wave was ignited by the *ab* initio theoretical prediction that HfTe$_5$ and ZrTe$_5$ crystals are located near the phase boundary between weak and strong TIs [140]. Li *et al.* investigated the ZrTe$_5$ single crystals by the ARPES technique, which revealed that the electronic structure of ZrTe$_5$ was consistent with a 3D DSM [141]. Moreover, obvious negative longitudinal MR was observed in ZrTe$_5$ when the magnetic and electrical fields were parallel. The NMR was suggested to arise from the chiral anomaly in the DSM ZrTe$_5$ [141], and the discovery further stimulated intensive studies [142-163]. Chen *et al.* reported the linear rising of the optical conductivity in a relatively large energy scale expected for a 3D DSM [142]. They also presented a magnetoinfrared spectroscopy study, in which the transitions between Landau levels and their field dependence were quantitatively consistent with the 3D massless Dirac fermions [143]. Besides, it was found the QL regime can be realized by a very small magnetic field (about 1 T) in the ZrTe$_5$ samples. Moreover, the observation of the splitting of the Landau levels was suggested as a transition in ZrTe$_5$ from a DSM to a NLSM [143].

The transport properties of ZrTe$_5$ were then provided by a lot of research work [144-160]. In ZrTe$_5$ nanoribbons exfoliated from bulk crystals, Zheng *et al.* presented exotic properties including the NMR at B//I, prominent SdH oscillations revealing a nontrivial π Berry phase, highly anisotropic effective masses and the ultrahigh mobility, which together signaled the existence of a 3D DSM phase in the ZrTe$_5$ [144]. Yuan *et al.* showed the quasi-2D Fermi surface and the bulk quantum Hall effect in ZrTe$_5$ [145]. Moreover, interesting transport phenomena in the ultraquantum regime of ZrTe$_5$ were reported [146-148]. In an extremely large magnetic field, Liu *et al.* observed sharp resistivity peaks in the ultraquantum limit [146]. The anomalous changes of the resistivity were attributed to the field-induced phase transitions and the interaction-induced spontaneous mass generation of the Dirac fermions. Wang *et al.* studied the ZrTe$_5$ samples with very low carrier density, in which a new class of quantum oscillations with logB periodicity was discovered beyond the QL [147]. The peculiar log-periodic MR oscillations manifest the rarely observed discrete scale invariance behavior in a condensed matter system, which was suggested to be closely related to the two-body quasi-bound states formed by massless Dirac fermions with Coulomb attraction and the long pursued atomic collapse phenomenon in topological systems. The work also provides a new perspective on the ground state of topological materials beyond QL.

The pressure-tuned properties of ZrTe$_5$ were also investigated. Superconductivity was induced in the DSM ZrTe$_5$ under the high pressure generated by a screw-pressure-type diamond anvil cell [149]. The superconductivity appears above 6.2 GPa with the critical temperature Tc increasing with applied pressure and reaching a maximum of 4.0 K at 14.6 GPa. At pressures above 21.2 GPa, the phase coexists with a second superconducting phase with the maximum Tc of about 6.0 K. Furthermore, it was found the observed two-stage superconducting behavior closely related to the structural phase transition under the pressure. The hydrostatic pressure shows different influence on the magnetotransport properties of ZrTe$_5$ [150]. With increasing pressure, the magnetoresistance in the case of B//b axis decreases drastically and the quantum oscillations reveals a nontrivial to trivial Berry phase transition. The results indicate the disruption of the DSM state under pressure. It was also observed that the ZrTe$_5$ evolves from a highly anisotropic to a nearly isotropic electronic system under the pressure, which suggests the quasi-2D nature of the Dirac cones in the material.

Though intensive experimental investigations support the DSM state in ZrTe$_5$, the evidences for identifying ZrTe$_5$ as a weak TI or a strong TI were also reported [160-165]. For example, the scanning tunneling microscopy suggested that the ZrTe$_5$ crystal as a weak 3D TI [160,161]. In fact, based on the recent intensive research from both theorists and experimentalists, it is known the ZrTe$_5$ material is extremely sensitive to the cell volume [166] and thus the measured physical properties, such as the resistance-temperature characteristic, the MR effect and the Hall information are divergent [144-160]. Impurities/nonstoichiometry, defects, and strain may affect the cell volumes, which are closely related to the growth condition of the samples. Thus, the Zr(Hf)Te$_5$ can manifest as a strong TI, a DSM, or a weak TI under variant crystal volumes, which depends on the sample quality [166]. Useful information for understanding the divergent bulk properties can be obtained from the investigations on ZrTe$_5$ nanosheets with different thickness [152,153], which also indicate a very sensitive band structure of ZrTe$_5$.

As a sister compound of ZrTe$_5$, HfTe$_5$ has been rarely studied experimentally. Wang *et al.* presented the first electrical transport evidence for the chiral anomaly and the ultrahigh mobility in HfTe$_5$ crystals (Fig. 7) [174]. In specific, anomalous NMR is observed when the magnetic field B is aligned along the direction of electrical field E. The NMR is quite sensitive to the orientation of B relative to E and disappears rapidly when B deviates from E. Quantitative analyses by the semi-classical formula $\sigma_{xx} = (\sigma_0 - a\sqrt{B})(1 + C_w B^2)$ suggest an origin from the chiral anomaly (Fig. 7(a)). The ultrahigh mobility and ultralow carrier density were revealed by analyzing the Hall traces (Fig.

7(b)). The mobility at 2 K, μ=1.84±0.01×$10^6$ $cm^2$/V/s, is comparable to that of Dirac fermions in $Cd_3As_2$ [33,35] and larger than that of $ZrTe_5$ [144,146]. Even at the room temperature, the mobility of $HfTe_5$ crystal is as high as 3700 $cm^2$/V/s, which is significant for the potential electronic applications. Similar results in $HfTe_5$ were also reported by Zhao *et al.* [175]. Furthermore, the superconductivity was observed in $HfTe_5$ by high pressure experiments [176,177].

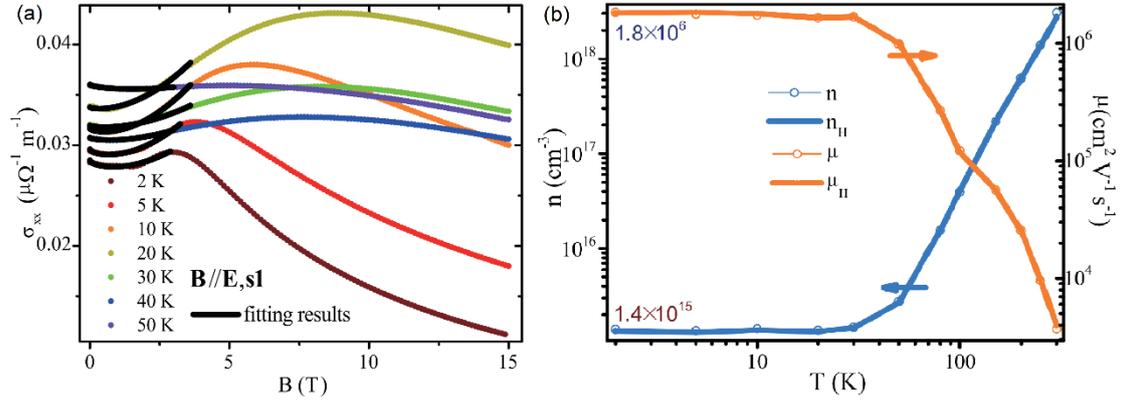

Fig. 7. (color online) (a) Quantitative fitting for the NMR in $HfTe_5$ when B//E by chiral anomaly model. Colorful dots are experimental data and black lines are fitting results. (b) Temperature dependence of the estimated carrier density and mobility in $HfTe_5$. [174]

Beyond the above DSM systems with normal Dirac nodes in the momentum space, the type II DSM $PtTe_2$ with titled Dirac nodes was also identified recently [178]. Yan *et al.* studied the bulk $PtTe_2$ single crystal and revealed a pair of strongly tilted Dirac cones by the combined ARPES measurements and first-principles calculations, which confirmed $PtTe_2$ as a DSM with the Lorentz-violating type-II Weyl/Dirac fermions [179]. The work paves way for further investigating novel quantum phenomena and topological phase transition in the new TSM sates.

## 3. Transport properties of Weyl semimetals
### 3.1 Chiral anomaly, ultrahigh mobility, and giant magnetoresistance

In WSMs, the Weyl nodes with different chirality are separated in momentum space. In parallel electric and magnetic fields (E||B), charge is predicted to flow between the Weyl nodes. This process violates the conservation of chiral charges and leads to an axial charge current. The axial current gives rise to better conductivity of the WSM and leads to a NMR phenomenon. This is the chiral (Adler-Bell-Jackiw) anomaly effect, which virtually is a remarkable phenomenon associated with the chiral Weyl fermions [110-115]. It is worth noting, however, the NMR under E∥B itself may not be a compelling signature of the chiral anomaly, especially when the observation is in a material where the chirality is not well defined [58,87,179,180]. Arnold *et al.* studied the WSM TaP when the chirality of the quasiparticles at the Fermi level is ill-defined, and a large NMR was observed [58]. It was found that a magnetic field dependent inhomogeneous current distribution is important in the sample, which indicates the NMR may be related to the current jet effect. Thus, in the experimental investigations, special care is needed to clarify if the NMR is from chiral anomaly.

The first discovery of E∥B sensitive NMR in topological materials was reported by Wang *et al.* in 2012 [181]. In topological insulator films, NMR was observed when E∥B, which disappears when the electric and magnetic fields are not parallel. Further control experiments revealed that the NMR is not related to the crystal orientations. A recent theoretical work suggested that the anomalous NMR in TI films could be understood quantitatively by a semiclassical formula, which considers the Berry curvature and orbital moment corrections but without chiral anomaly [180]. Thus, the Berry curvature might be a more general origin for the E∥B sensitive NMR in topological materials.

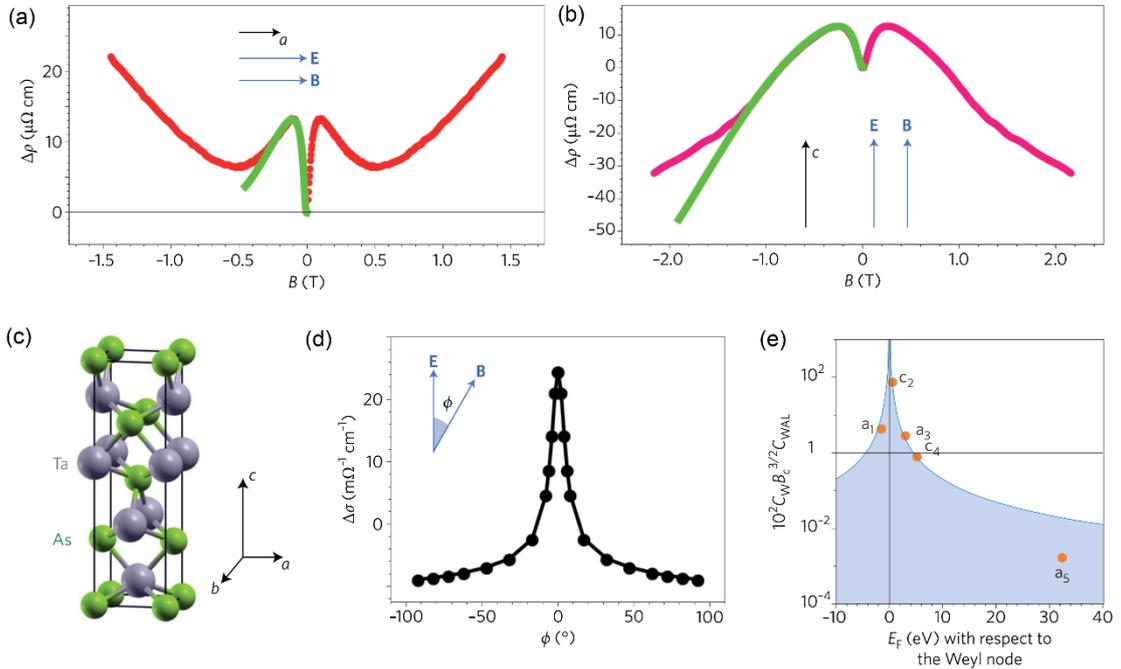

Fig. 8. (color online) (a,b) Negative MR in TaAs. Red lines are experimental data and green lines are theoretical fits. a and c are the lattice vectors. (c) The tetragonal crystal lattice of TaAs. (d) MR data as a function of the angle between the electric and magnetic fields. (e) Dependence of the chiral coefficient on chemical potential $E_F$. [86]

The clear identification of Weyl nodes in TaAs crystals paves the way for the exploration of chiral anomaly. Huang *et al*. reported an NMR in the characterized WSM TaAs when the external magnetic and electrical fields are parallel [61]. The NMR shows sharp dependence on the angle between the electric and magnetic fields and can be well fitted by the chiral anomaly formula, which was recognized as the signature of the expected chiral anomaly. At almost the same time, Zhang *et al*., presented the evidence of chiral anomaly in TaAs by showing the unique features of the NMR (Fig. 8), which includes (1) the dependence on the angle between the electric and magnetic fields; (2) the independence on the direction of the field with respect to the crystalline axis; (3) the relation to the chemical potential $E_F$. Moreover, the transport data are corroborated by photoemission measurements, first-principles calculations and theoretical analyses, which collectively demonstrated the existence of the chiral anomaly in TaAs driven by Weyl fermions [59,86].

The magnetotransport measurements also demonstrate other intriguing properties of WSM TaAs, such as the giant MR embedding with obvious SdH oscillations, nontrivial Berry phase, and ultrahigh mobility. Zhang *et al.* reported that the bulk TaAs exhibits an extremely large, unsaturated linear MR with a magnetic field perpendicular to the current [60]. The MR reaches 540000 % at 10 K in a magnetic field of 9 T and $2.47\times10^6$ % at 1.5 K in a magnetic field of 56 T. Besides, the Landau fan diagram of the superimposed SdH oscillations on the MR shows the nontrivial Berry phase close to π. Due to the semimetal property of TaAs, both hole and electron carriers exist in the system, which usually induces the observation of nonlinear Hall behavior. The nonlinear Hall trace can be analyzed by a two-carrier model to reveal the carrier information. Zhang *et al.* showed that TaAs manifests an ultrahigh carrier mobility ($5\times10^5 cm^2/V\cdot s$ at 2 K) [60], which is consistent with the result ($1.8\times10^5 cm^2/V\cdot s$ at 10 K) reported by Huang *et al.* [61]. The sample dependence of the transport properties was further investigated by Zhang *et al.* in details [60]. It was found that the growth procedure significantly affects the sample's quality, which is crucial to the transport properties such as the MR effect and mobility. Particularly, the position of the Fermi level strongly influences the scattering process and mobility, which indicates it is important to fine tune the Fermi level for exploring the unique properties of the Weyl quasiparticles.

3.2 Ultraquantum features

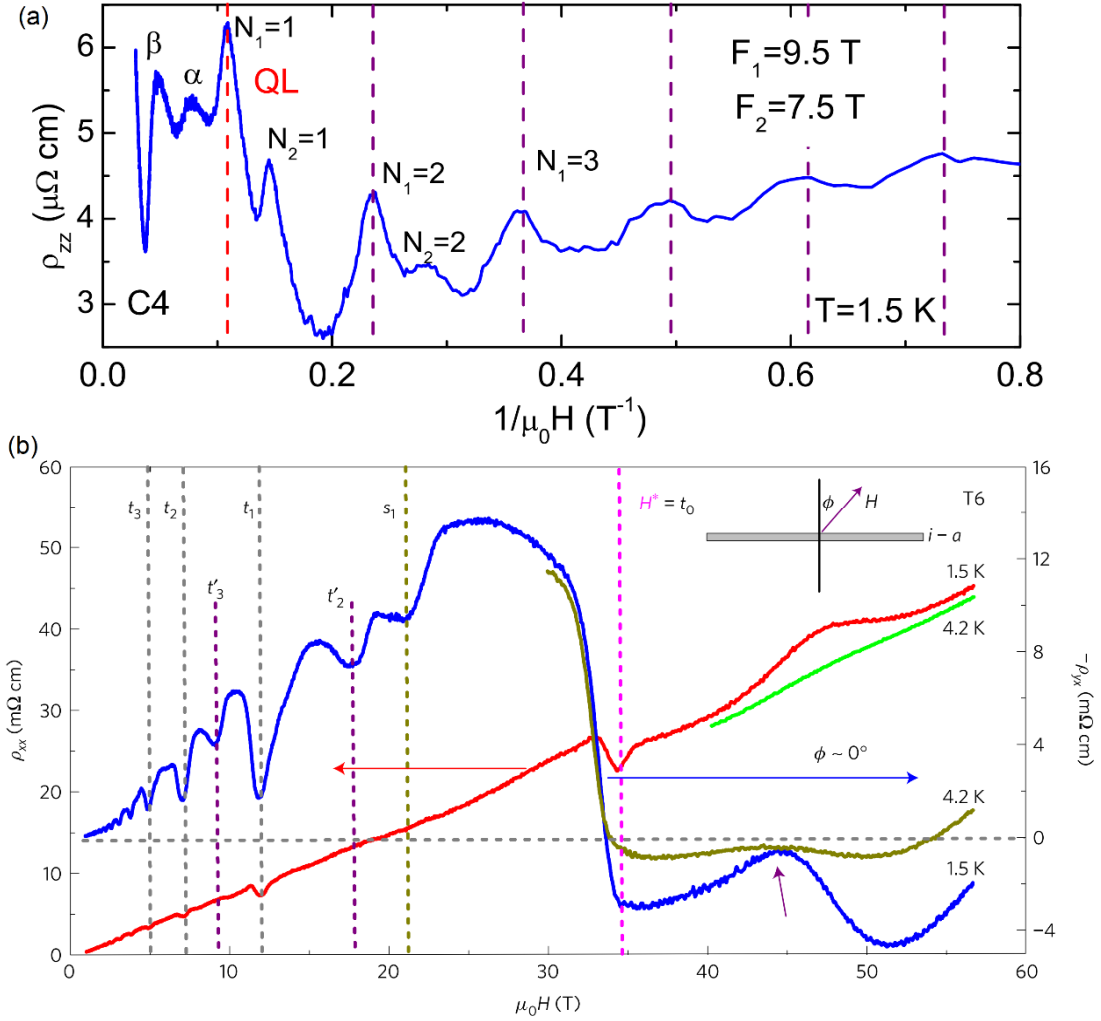

Fig. 9. (color online) (a) The longitudinal MR of TaAs with respect to the inverse field at 1.5 K [182]. (b) The MR and Hall of TaP in ultrahigh magnetic fields [186]. Anomalous features occur in the ultraquantum regime of the WSMs above the QL.

Since the FSs of WSMs usually are small, it is achievable to reach the QL in available magnetic fields [182-186]. When the electrons are driven into the lowest Landau level in an intensive magnetic field, anomalous features in the ultraquantum regime were reported in WSMs [182-186]. Zhang *et al.* reported the electrical transport properties of WSM TaAs in magnetic fields up to 55 T [182]. Additional oscillatory structures were observed in the longitudinal MR and Hall traces beyond the QL of TaAs at low temperatures (Fig. 9). According to the temperature dependent behaviors of the structures, the origin was suggested to arise from the instability of Weyl electrons at high magnetic fields [182]. The longitudinal NMR in TaAs was studied across the QL regime with a maximum field of 45 T, and it was found to disappear when the QL was reached [183]. The observations excluded the current jetting effect as origin of the NMR and confirmed the intrinsic NMR to be associated with the Weyl fermions. More importantly, a hysteretic phase-transition was discovered in TaAs upon approaching and surpassing the QL [183]. Ramshaw *et al.* used ultrahigh magnetic fields up to 95 T to explore new states in the WSM TaAs by driving it into the ultraquantum regime [184]. The exotic experimental results indicated some interesting phenomena, such as the gapping of the lowest chiral Landau level and a field-induced phase transition.

In topologically trivial conductors, the magnetization tends to a saturation at the QL, which was commonly observed in low carrier density semiconductors e.g. InSb and InAs [187]. In WSMs, however, the collapse of the magnetization leads to a magnetic anomaly [185]. In the magnetic torque measurements on WSM NbAs, a sharp anomaly, the sign change of the magnetic torque, was reported upon reaching the QL in high magnetic fields. The observation of crossover from diamagnet below the QL to paramagnet above the QL was attributed to the topological nature of the Weyl electrons [185].

Upon approaching and surpassing the QL, the WSM TaP also presents a series of anomalous structures in the magnetic torque and electrical transport measurements, which were ascribed to the multiple field-induced phase transitions [183]. Zhang *et al.* studied the magnetic tunneling of the lowest Landau bands in TaP [186], which indicates the hybridization of adjacent Fermi pockets. To observe the magnetic tunneling of the lowest Landau level, the QL should be reached and the separation of pockets in k space cannot be far. In the high-field measurements, Zhang *et al.* observed a sharp sign reversal in the Hall resistivity at a magnetic field located inside the ultraquantum limit. The magnetic field corresponds to the separation of the W1 Weyl nodes in momentum space, which reveals the magnetic coupling between the electron pockets arising from the W1 Weyl fermions. The findings in the ultraquantum regime provide evidences for the magnetic-field-induced gap opening of the chiral Landau levels and the unique Weyl node annihilation in the WSM [186].

The fertile phenomena observed in the ultraquantum WSMs shows that the QL is a promising regime for finding new states of matter based on the Weyl fermions [188]. Thus, the related investigations in future may bring more interesting discoveries in this direction. On the other hand, though possible mechanisms of features in the ultraquantum regime of WSMs have been argued, further investigations are still necessary to conclusively identify the exact underlying nature.

3.3 Superconductivity in Weyl semimetals

As introduced in Section 2.5, the TSC is a charming phase of matter due to the importance in physics and potential applications for topological quantum computation. Superconductivity results from attractive electron-electron interactions, which may arise at low temperatures in almost any metal in theory. For a WSM with Weyl fermion and topological FA surface states, it could be a natural candidate for the realization of topologically nontrivial superconductors if superconductivity can be induced. Thus, it is crucial to explore the possible superconductivity in WSMs. Despite growing theoretical interests in this topic, the experimental realization of superconductivity in WSMs is quite challenging.

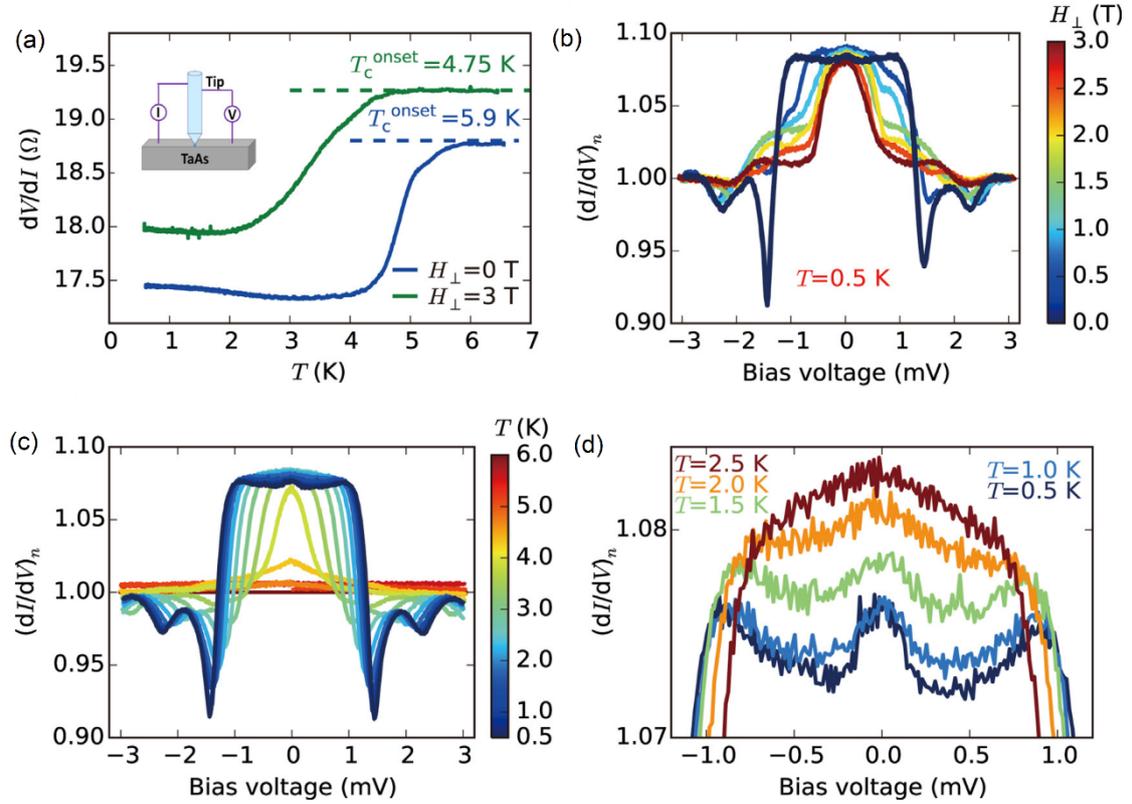

Fig. 10. (color online) (a) Temperature dependence of PC resistance showing superconducting signal. Inset: schematics of the PCS measurement configuration. (b) Normalized dI/dV spectra at 0.5 K under out-of-plane magnetic fields. The navy-blue line denotes zero field measurement. (c) Normalized dI/dV spectra at selected temperatures from 0.5 K to 6.0 K. (d) Zoom in of the normalized ZBCP. [189]

In 2016, Wang *et al.* reported unconventional superconductivity induced by hard PC on 3D DSM $Cd_3As_2$ crystals (more details in Section 2.5) [131]. The results reveal a new way to detect and study potential topological superconductivity by using hard tip/PC on topological materials. Soon after that, Wang *et al.* further discovered tip-induced superconductivity by hard PC on WSM TaAs crystals for the first time, which might be topologically nontrivial (Fig. 10) [189]. They used high quality TaAs that has been characterized as WSMs. The electrical transport properties at ambient environment show obvious metallic behavior of the sample. PC measurements were then carried out between a mechanically sharpened PtIr tip and the (001) surface of TaAs single crystal. A significant resistance drop was observed in the RT behavior, suggesting existence of tip-induced superconductivity in the TaAs crystal. The suppression of $T_c$ under a magnetic field is also consistent with the property of superconductivity (Fig. 10(a)). The PCS at 0.5 K show a small ZBCP and a wide conductance plateau with double peaks ended with sharp dips symmetric to zero bias (Fig. 10(b)). Further study on the PCS at selected perpendicular magnetic fields and the temperature dependence of the PCS reveal the unconventional characteristic of the superconductivity in the WSM TaAs (Figs. 10(b)-10(d)). Theoretical analyses on the experimental results show that the induced superconductivity on TaAs may have nontrivial topology protected by a mirror-symmetry in TaAs. Therefore, this work demonstrates that superconducting TaAs is a TSC candidate.

The hard PC results of the WSM TaAs indicate that the local "uniaxial" pressure and doping effect around PC region may be important to the appearance of superconductivity. The high-pressure experiment on TaAs was also reported [190]. At a pressure around 14 GPa, it was predicted that TaAs might show a new WSM phase, which has 12 WPs at the same energy level rather than the 24 WPs distributed at two energy levels [190]. Experimental results consistent with the new hexagonal P-6m2 phase at pressure were provided by the electrical transport measurements and Raman spectroscopy. Besides, the P-6m2 phase was suggested to be kept to ambient pressure. However, no hint of superconductivity appears down to 1.8 K until 54.0 GPa in this high-pressure work [190].

By doing resistance measurements on the WSM TaP under pressure up to about 100 GPa, Li *et al.* observed the superconductivity in TaP at about 70 GPa [64]. The superconductivity retains when the pressure is released. Their calculations based on the density functional theory illustrate a structure transition of TaP at about 70GPa, which may be the underlying origin of superconductivity. The discovery of superconductivity in WSMs TaAs and TaP mentioned above will stimulate further study on the superconductivity in WSMs and offer a new platform to search for TSCs. Especially for the tip-induced superconductivity on TaAs, it might be interface superconductivity and the FA surface states may contribute a lot since both tip and sample are non-superconducting and the surface state carrier is easier to be modulated by tip to form superconducting Cooper pairs. Further theoretical and experimental investigations are highly desirable to classify this new topic.

3.4 Beyond the TaAs family

TaAs family are the most studied WSMs while the TaAs family are only on behalf of one type of WSMs, known as type-I WSMs. WSMs can be further classified into type-I WSMs and type-II WSMs based on whether the Weyl cones are titled or not (Fig. 11). For the type-I WSM, the two bands touch at separate WPs in the momentum space (Fig. 11(a)). For the type-II WSMs, the Weyl cones are titled and the bands that touch at the Weyl point now overlap in energy, forming electron and hole pockets (Fig. 11(b)) [191]. The Weyl fermion in type-II WSMs manifests Lorentz-violating behavior.

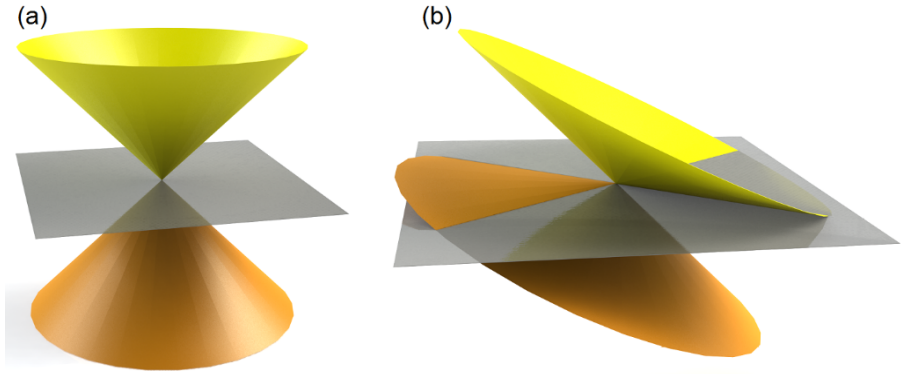

Fig. 11. (color online) (a) The Weyl cone in Type-I WSMs. (b) The Weyl cone in Type-II WSMs. The grey planes correspond to the position of the Fermi level.

Theoretically predicted type-II WSM $MoTe_2$ was recently confirmed by showing direct spectroscopic evidence of topological FAs in the ARPES results [192-199]. The Type-II WSM $MoTe_2$ is found to be superconducting around 0.1 K [200] and the doped $MoTe_2$ can show higher Tc [201]. Li *et al.* grew $MoTe_2$ crystals by partially substituting Te with S to enhance the superconducting onset temperature and observed nontrivial superconducting signatures in this topological material [202]. By scanning tunneling spectroscopy measurement, they firstly observed the quasiparticle interference patterns of FAs at the surface of $MoTe_{2-x}S_x$ crystal, which are consistent with the calculated nontrivial band structure. They also detected a relatively large superconducting gap ($\Delta$) on the sample surface, which shows a large $\Delta/k_BT_c$ value 8.6. This value is much larger than that of the conventional weak-coupling superconductors, and also larger than the bulk superconducting gaps fitted from the specific heat measurement. The large superconducting gap might be from superconductivity parity mixing or unconventional paring mechanism on the surface, potentially indicating the nontrivial superconductivity from FA surface states. Their further transport measurements (Fig. 12) show two-band superconductivity with potential s+- pairing from a dominant interband coupling, which is consistent with the specific heat results. The s+- pairing was ever conjectured for undoped $MoTe_2$ under high pressure due to the fact that the superconductivity of $MoTe_2$ is sensitive to disorders [203]. The large surface superconducting gap and s+- paring superconductivity make the WSM $MoTe_{2-x}S_x$ a promising TSC candidate [202].

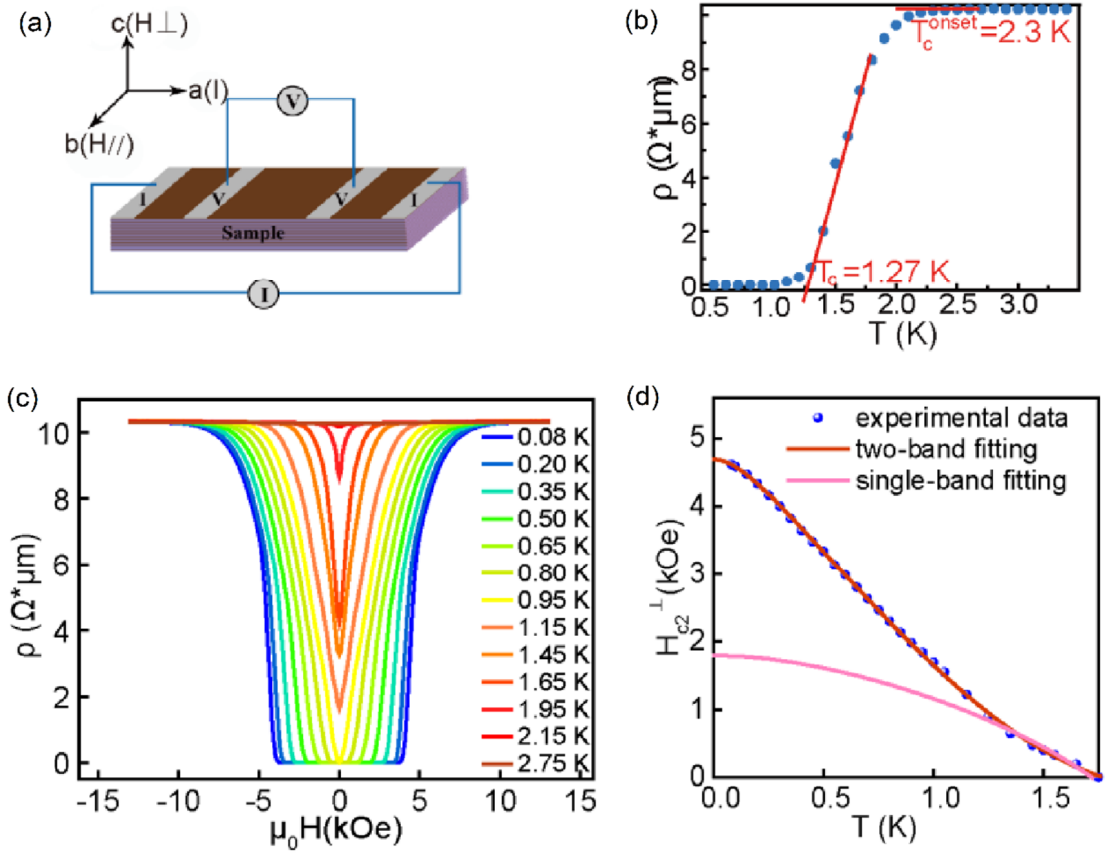

Fig. 12. (color online) (a) Schematic diagram for magnetotransport measurements of $MoTe_{2-x}S_x$. (b) Typical resistivity–temperature curve of $MoTe_{2-x}S_x$ showing superconductivity. (c) MR at various temperatures under perpendicular field. (d) Temperature dependence of $H_{c2}$ extracted from the data in (c). The red curve is the best fit of two-band s-wave model to the experimental data. The fitting parameters showing strong interband coupling, indicates s+- superconductivity in $MoTe_{2-x}S_x$. [202]

WTe$_2$, a semimetal which was studied intensively in the past years due to the large MR [204-211], was also predicted to be a Type-II WSM [191]. Spectroscopic evidence on the type II WSM WTe$_2$ has been provided by ARPES experiments while the FAs are hard to be clearly resolved [212]. Moreover, significant anisotropic chiral anomaly effect of the type-II WSM has been observed in the WTe$_2$ thin films and the Fermi-level delicately adjusted WTe$_{1.98}$ crystals by performing systematic magnetotransport studies [213,214].

An additional subclass of WSM materials is the so-called multi-WSMs [215]. The monopole charge of a Weyl point is equal to the number of Berry fluxes passing through a closed surface enclosing the target Weyl point only. So far, most studies on WSMs have focused on the case with the lowest monopole charge, C = ±1, such as the TaAs family. The Weyl point in multi-WSMs carries topological charges of higher magnitude, for example, HgCrSe$_2$ and SrSi$_2$ are the so-called double WSM candidates with C = ±2 predicted by the calculations [216,217]. Such higher-charge Weyl points may be stabilized by certain point group crystal symmetries, and more interesting effect may appear. For instance, it is pointed out by theory that WPs with higher monopole charge exhibit anisotropic screening to the Coulomb interactions [218,219].

## 4. Perspective on the study of topological semimetals

In this chapter, we have mainly introduced the transport properties of DSMs and WSMs in magnetic fields. It is noted that the TSMs, either WSMs or DSMs, show many features in common, such as the large MR, the ultrahigh mobility, and the remarkable quantum oscillations. On the other hand, we can see that each TSM system also possesses its unique property, such as the nested Fermi surface of Cd$_3$As$_2$ and the low QL of TaAs. In this view, the many other predicted materials still deserve experimental identifications and more intriguing phenomena can be expected in those new TSM materials [11]. In addition, the exploration of new type of TSMs is also a significant issue in this field. As is known, the currently studied DSMs Cd$_3$As$_2$ and Na$_3$Bi possess a pair of DPs while the WSM TaAs family have 12 pairs of WPs. The large number of the DPs or WPs may hinder a direct investigation and understanding of the properties of TSMs. Thus, the TSMs with the simplest WPs are desirable. Besides, for the precise demonstrations of the characteristic of Weyl fermions, the TSMs with ideal Fermi level very close to the nodes and no other states at it are highly expected. So far, most investigated WSM materials are inversion symmetry broken systems, the search for magnetic WSM materials with broken TRS is very important direction too [220]. It is noted that signature of the existence of Weyl fermions has been reported in magnetic Heusler GdPtBi [221]. The great diversity of Heusler compounds with more than 1500 members may establish an opportunity to enlarge the TSM family [222]. Besides, the recent discovery of magnetic topological semimetal states in $Sr_{1-y}Mn_{1-z}Sb_2$ (y, z < 0.1) and NdSb is also of interest [223,224].

In fact, both Dirac and Weyl fermions can be considered as special cases of the "multifold fermions" [225]. Due to the two/four-fold degeneracies at high-symmetry points, Weyl/Dirac quasiparticles are thus two/four-fold fermions within the multifold-fermion scenario. In solid-state systems with spin-orbit coupling and with certain symmetries, new types of symmetry–protected fermions arise by possessing three-, six- and eight-fold degeneracies at and along high-symmetry points, lines and surfaces. In analogous to Dirac/Weyl semimetals, emergent phenomena, e.g., Weyl points, FAs, Fermi drums and chiral anomaly, can be expected in the multifold-fermion systems. First-principles calculations have predicted several candidate materials that might host the exotic multifold-fermionic quasiparticles [225-228], e.g., $Ag_3AuSe_2$, $Ba_4Bi_3$, MgPt, $CuBi_2O_4$, θ-TaN, WC-type ZrTe, etc. Very recently, the scientists from IOP china have reported the ARPES experimental signatures of the triply degenerate point in the electronic structure of crystalline MoP and WC [229,230]. Lv *et al.* observed the three-component fermions in the bulk of MoP, which is well consistent with the theoretical results [229]. Moreover, the triply degenerate points were also discovered by Ma *et al.* in high quality WC, in which the nodal points were near the Fermi level and the exotic properties related to the three-component fermions could be detected more clearly [230]. Importantly, they further observed FA topological surface states connecting to the surface projections of the triply degenerate nodal points, which suggested nontrivial topology of the semimetal state [230]. In addition, the electrical transport properties of the new type of semimetals were investigated. He *et al.* found the NMR phenomenon in WC can only appear for certain direction when the magnetic field was parallel to the current [231], and Shekhar *et al.* reported extremely low resistivity in MoP [232], which may be related to the exotic triple point fermions.

As introduced in the above, topological superconductivity is a hot topic in condensed matter physics. Previous experimental explorations are mainly focused on TI (spin-orbit coupling semiconductor)-superconductor hybrid structures. The discovery of TSMs offers new and ideal platforms to study the topological superconductivity due to the naturally nontrivial topology. Particularly, the superconductivity has been observed in some DSMs and WSMs, while the topology of the superconductivity still needs more theoretical and experimental efforts to fully illustrate the issue. Moreover, the hard PC method has been developed as a new and effective way to induce superconductivity on non-superconducting TSMs [131], which significantly expands the research subjects. We expect that the topological superconductivity could be finally confirmed by this new method combing with other experimental techniques and pave the way for future topological quantum computing.

As the Fermi levels of type-I WSMs and DSMs are at or near the discrete Weyl nodes, it is expected that these systems would usually show small carrier density and small QL. For example, the QL of TaAs can be about 8 T [60] and the QL of $ZrTe_5$ can be below 1 T [143,147]. Thus, the ultraquantum regime has now been an achievable territory to be exploited. For example, anomalous features were reported in the ultraquantum regime of WSM TaAs and TaP. Though experimental results are obvious and distinct, the full understanding of the exotic phenomena still needs more investigations from theorists and experimentalists in the future. Moreover, the very small carrier density system may be recognized as an electron gas system and the few-body and many-body effect would be rather relevant. Thus, more interesting phenomena are expected to be revealed in the ultraquantum regime of TSMs.

TSMs not only show unique properties in the electrical transport behaviors but also bring surprising phenomena in the optical and thermal measurements. For example, DSM $Cd_3As_2$ not only processes advantages of graphene as a photosensitive material but also shows great potential to realize either an ultrafast broadband photodetector or an ideal ultrafast switch in mid-infrared lasers and photonics [233-235]. Besides, a negative magneto-thermopower possibly induced by the chiral anomaly is observed in thermopower measurements when the magnetic field is parallel to the temperature gradient and the quadratic coefficient is found nearly twice of that for the electrical conductivity [236]. Furthermore, the electron or hole doping in DSM $Cd_3As_2$ largely improves the optimum figure of merit, which shows a great potential for the high performance of thermoelectric applications [237].

The experimental confirmations of the DSM and WSM were initially realized in 2014 and 2015, respectively. Though great progress has been made in this field, in fact, the current investigations are only the tip of an iceberg. More classes of DSMs, WSMs, and other types of TSMs are waiting to be uncovered. Further efforts are needed to be devoted to discovering new alluring phenomena in these unique systems and to revealing the underlying physics, which would pave the way for the understanding and potential applications of the materials in topological world. Significant breakthroughs are expected to be generated in this rapidly developing field, wherein the application of high magnetic fields definitely is one of the most powerful methods.

## Acknowledgements

We acknowledge Yanzhao Liu for the help in the preparation of this book chapter. This work was financially supported by the National Basic Research Program of China (Grant No. 2018YFA0305604; Grant No. 2017YFA0303300; Grant No. 2013CB934600), the Research Fund for the Doctoral Program of Higher Education (RFDP) of China (20130001110003), the Open Project Program of the Pulsed High Magnetic Field Facility (Grant No. PHMFF2015002) at the Huazhong University of Science and Technology, the Open Research Fund Program of the State Key Laboratory of Low-Dimensional Quantum Physics, Tsinghua University under Grant No. KF201703, and the Key Research Program of the Chinese Academy of Sciences (Grant No. XDPB08-2).

.

## References

[1]     Hasan M Z and Kane C L 2010 *Rev. Mod. Phys.* **82** 3045
[2]     Qi X L and Zhang S C 2011 *Rev. Mod. Phys.* **83** 1057
[3]     Moore J E 2010 *Nature* **464** 194
[4]     Qi X L and Zhang S C 2010 *Physics Today* **63** 33
[5]     Yan B and Zhang S C 2012 *Rep. Prog. Phys.* **75** 096501
[6]     Ando Y 2013 *J Phys. Soc. Jpn.* **82** 102001
[7]     Weng H, Dai X and Fang Z 2014 *MRS Bull.* **39** 849
[8]     Hasan M Z, Xu S Y and Bian G 2015 *Phys. Scr.* **2015** 014001
[9]     Wan X, Turner A M, Vishwanath A and Savrasov S Y 2011 *Phys. Rev. B* **83** 205101
[10]    Burkov A A and Balents L 2011 *Phys. Rev. Lett.* **107** 127205
[11]    Weng H, Dai X and Fang Z 2016 *J. Phys. Condens. Matter* **28** 303001
[12]    Armitage N P, Mele E J and Vishwanath A 2018 *Rev. Mod. Phys.* **90** 15001
[13]    Yan B and Felser C 2017 *Annu. Rev. Conden. Ma. P.* **8** 337
[14]    Young S M, Zaheer S, Teo J C Y, Kane C L, Mele E J and Rappe A M 2012

*Phys. Rev. Lett.* **108** 140405

[15] Wang Z, Sun Y, Chen X-Q, Franchini C, Xu G, Weng H, Dai X and Fang Z 2012 *Phys. Rev. B* **85** 195320

[16] Wang Z, Weng H, Wu Q, Dai X and Fang Z 2013 *Phys. Rev. B* **88** 125427

[17] Liu Z K, Zhou B, Zhang Y, Wang Z J, Weng H M, Prabhakaran D, Mo S K, Shen Z X, Fang Z, Dai X, Hussain Z and Chen Y L 2014 *Science* **343** 864

[18] Kim H J, Kim K S, Wang J F, Sasaki M, Satoh N, Ohnishi A, Kitaura M, Yang M and Li L 2013 *Phys. Rev. Lett.* **111** 246603

[19] Sato T, Sato T, Segawa K, Kosaka K, Souma S, Nakayama K, Eto K, Minami T, Ando Y and Takahashi T 2011 *Nat. Phys.* **7** 840

[20] Singh B, Sharma A, Lin H, Hasan M Z, Prasad R and Bansil A. 2012 *Phys. Rev. B* **86** 115208

[21] Novak M, Sasaki S, Segawa K and Ando Y 2015 *Phys. Rev. B* **91** 041203

[22] Orlita M, Basko D M, Zholudev M S, Teppe F, Knap W, Gavrilenko V I, Mikhailov N N, Dvoretskii S A, Neugebauer P, Faugeras C, Barra A L, Martinez G and Potemski M 2014 *Nat. Phys.* **10** 233

[23] Brahlek M, Bansal N, Koirala N, Xu S Y, Neupane M, Liu C, Hasan MZ and Oh S. 2012 *Phys. Rev. Lett.* **109** 186403

[24] Liu J and Vanderbilt D 2013 *Phys. Rev. B* **88** 224202

[25] Wu L, Brahlek M, Valdés Aguilar R, Stier A V, Morris C M, Lubashevsky Y, Bilbro L S, Bansal N, Oh S and Armitage N P 2013 *Nat. Phys.* **9** 410

[26] Dziawa P, Kowalski B J, Dybko K, Buczko R, Szczerbakow A, Szot M, Łusakowska E, Balasubramanian T, Wojek B M, Berntsen M H, Tjernberg O and Story T 2012 *Nat. Mater.* **11** 1023

[27] Neupane M, Xu S Y, Sankar R, Alidoust N, Bian G, Liu C, Belopolski I, Chang T R, Jeng H T, Lin H, Bansil A, Chou F and Hasan M Z 2014 *Nat. Commun.* **5** 3786

[28] Liu Z K, Jiang J, Zhou B, Wang Z J, Zhang Y, Weng H M, Prabhakaran D, Mo S K, Peng H, Dudin P, Kim T, Hoesch M, Fang Z, Dai X, Shen Z X, Feng D L, Hussain Z and Chen Y L 2014 *Nat. Mater.* **13** 677

[29] Borisenko S, Gibson Q, Evtushinsky D, Zabolotnyy V, Büchner B and Cava R J 2014 *Phys. Rev. Lett.* **113** 027603

[30] Yi H, Wang Z, Chen C et al. 2014 *Sci. Rep.* **4** 6106

[31] Xiong J, Kushwaha S K, Liang T, Krizan J W, Hirschberger M, Wang W, Cava

R J and Ong N P 2015 *Science* **350** 413

[32] Jeon S, Zhou B B, Gyenis A, Feldman B E, Kimchi I, Potter A C, Gibson Q D, Cava R J, Vishwanath A and Yazdani A 2014 *Nat. Mater.* **13** 851

[33] Liang T, Gibson Q, Ali M N, Liu M, Cava R J and Ong N P 2015 *Nat. Mater.* **14** 280

[34] He L P, Hong X C, Dong J K, Pan J, Zhang Z, Zhang J and Li S Y 2014 *Phys. Rev. Lett.* **113** 246402

[35] Zhao Y F, Liu H W, Zhang C L, Wang H C, Wang J F, Liu Z Q, Xing Y, Lu H, Liu J, Wang Y, Brombosz S M, Xiao Z L, Jia S, Xie X C and Wang J 2015 *Phys. Rev. X* **5** 031037

[36] Feng J, Pang Y, Wu D, Wang Z, Weng H, Li J, Dai X, Fang Z, Shi Y and Lu L 2015 *Phys. Rev. B* **92** 081306

[37] Steinberg J A, Young S M, Zaheer S, Kane C L, Mele E J and Rappe A M 2014 *Phys. Rev. Lett.* **112** 036403

[38] Du Y, Wan B, Wang D, Sheng L, Duan C G and Wan X 2015 *Sci. Rep.* **5** 14423

[39] Gibson Q D, Schoop L M, Muechler L, Xie L S, Hirschberger M, Ong N P, Car R and Cava R J 2015 *Phys. Rev. B* **91** 205128

[40] Yu R, Weng H, Fang Z, Dai X and Hu X 2015 *Phys. Rev. Lett.* **115** 036807

[41] Xu C Z, Chan Y H, Chen Y, Chen P, Wang X, Dejoie C, Wong M H, Hlevyack J A, Ryu H, Kee H Y, Tamura N, Chou M Y, Hussain Z, Mo S K and Chiang T C 2017 *Phys. Rev. Lett.* **118** 146402

[42] Herring C 1937 *Phys. Rev.* **52** 365

[43] Halász G B and Balents L 2012 *Phys. Rev. B* **85** 035103

[44] Bulmash D, Liu C X and Qi X L 2014 *Phys. Rev. B* **89** 081106R

[45] Liu J and Vanderbilt D 2014 *Phys. Rev. B* **90** 155316

[46] Weng H, Fang C, Fang Z, Bernevig B A and Dai X 2015 *Phys. Rev. X* **5** 011029

[47] Huang S M, Xu S Y, Belopolski I, Lee C C, Chang G, Wang B, Alidoust N, Bian G, Neupane M, Zhang C, Jia S, Bansil A, Lin H and Hasan M Z 2015 *Nat. Commun.* **6** 7373

[48] Lv B Q, Xu N, Weng H M, Ma J Z, Richard P, Huang X C, Zhao L X, Chen G F, Matt C E, Bisti F, Strocov V N, Mesot J, Fang Z, Dai X, Qian T, Shi M and Ding H 2015 *Nat. Phys.* **11** 724

[49] Xu S Y, Belopolski I, Alidoust N et al. 2015 *Science* **349** 613

[50] Lv B Q, Weng H M, Fu B B, Wang X P, Miao H, Ma J, Richard P, Huang X C,

Zhao L X, Chen G F, Fang Z, Dai X, Qian T and Ding H 2015 *Phys. Rev. X* **5** 031013

[51] Xu S Y, Belopolski I, Sanchez D S et al. 2015 *Sci. Adv.* **1** e1501092

[52] Xu S Y, Alidoust N, Belopolski I et al. 2015 *Nat. Phys.* **11** 748

[53] Yang L X, Liu Z K, Sun Y, Peng H, Yang H F, Zhang T, Zhou B, Zhang Y, Guo Y F, Rahn M, Prabhakaran D, Hussain Z, Mo S K, Felser C, Yan B and Chen Y L 2015 *Nat. Phys.* **11** 728

[54] Yang X, Liu Y, Wang Z, Zheng Y and Xu Z arXiv:1506.03190

[55] Ghimire N J, Luo Y, Neupane M, Williams D J, Bauer E D and Ronning F. 2015 *J. Phys. Condens. Matter* **27** 152201

[56] Hu J, Liu J Y, Graf D, Radmanesh S M, Adams D J, Chuang A, Wang Y, Chiorescu I, Wei J, Spinu L and Mao Z Q 2016 *Sci. Rep.* **6** 18674

[57] Du J H, Wang H D, Chen Q, Mao Q H, Khan R, Xu B J, Zhou Y X, Zhang Y N, Yang J H, Chen B, Feng C M, and Fang M H 2016 *Sci. China Phys. Mech. Astron.* **59** 657406

[58] Arnold F, Shekhar C, Wu S C, Sun Y, Dos Reis R D, Kumar N, Naumann M, Ajeesh M O, Schmidt M, Grushin A G, Bardarson J H, Baenitz M, Sokolov D, Borrmann H, Nicklas M, Felser C, Hassinger E and Yan B 2016 *Nat. Commun.* **7** 11615

[59] Zhang C L, Xu S Y, Belopolski I et al. 2016 *Nat. Commun.* **7** 10735

[60] Zhang C L, Yuan Z, Jiang Q D, Tong B, Zhang C, Xie X C and Jia S 2017 *Phys. Rev. B* **95** 085202

[61] Huang X, Zhao L, Long Y, Wang P, Chen D, Yang Z, Liang H, Xue M, Weng H, Fang Z, Dai X and Chen G 2015 *Phys. Rev. X* **5** 031023

[62] Zhang C, Guo C, Lu H, Zhang X, Yuan Z, Lin Z, Wang J and Jia S 2015 *Phys. Rev. B* **92** 041203

[63] Xu N, Weng H M, Lv B Q, Matt C E, Park J, Bisti F, Strocov V N, Gawryluk D, Pomjakushina E, Conder K, Plumb N C, Radovic M, Autes G, Yazyev O V, Fang Z, Dai X, Qian T, Mesot J, Ding H and Shi M 2016 *Nat. Commun.* **7** 11006

[64] Li Y F, Zhou Y H, Guo Z P, Han F, Chen X L, Lu P C, Wang X F, An C, Zhou Y, Xing J, Du G, Zhu X Y, Yang H, Sun J, Yang Z R, Yang W G, Mao H K, Zhang Y H and Wen H H 2017 *npj Quantum Materials* **2** 66

[65] Xu D F, Du Y P, Wang Z, Li Y P, Niu X H, Yao Q, Dudin P, Xu Z A, Wan X G and Feng D L 2015 *Chin. Phys. Lett.* **32** 107101

[66] Shekhar C, Nayak A K, Sun Y, Schmidt M, Nicklas M, Leermakers I, Zeitler U, Skourski Y, Wosnitza J, Liu Z, Chen Y, Schnelle W, Borrmann H, Grin Y, Felser C and Yan B 2015 *Nat. Phys.* **11** 645

[67] Souma S, Wang Z W, Kotaka H, Sato T, Nakayama K, Tanaka Y, Kimizuka H, Takahashi T, Yamauchi K, Oguchi T, Segawa K and Ando Y 2016 *Phys. Rev. B* **93** 161112

[68] Wang Z, Zheng Y, Shen Z X, Lu Y H, Fang H Y, Sheng F, Zhou Y, Yang X J, Li Y P, Feng C M, and Xu Z A 2016 *Phys. Rev. B* **93** 121112

[69] Fang C, Weng H, Dai X and Fang Z 2016 *Chin. Phys. B* **25** 117106

[70] Burkov A A, Hook M D and Balents L 2011 *Phys. Rev. B* **84** 2011

[71] Weng H, Liang Y, Xu Q, Yu R, Fang Z, Dai X and Kawazoe Y 2015 *Phys. Rev. B* **92** 045108

[72] Ali M N, Gibson Q D, Klimczuk T and Cava R J 2014 *Phys. Rev. B* **89** 020505

[73] Bian G, Chang T R, Sankar R, Xu S Y, Zheng H, Neupert T, Chiu C K, Huang S M, Chang G, Belopolski I, Sanchez D S, Neupane M, Alidoust N, Liu C, Wang B, Lee C C, Jeng H T, Zhang C, Yuan Z, Jia S, Bansil A, Chou F, Lin H and Hasan MZ 2016 *Nat. Commun.* **7** 10556

[74] Yamakage A, Yamakawa Y, Tanaka Y and Okamoto Y 2015 *J. Phys. Soc. Jpn.* **85** 013708

[75] Xie L S, Schoop L M, Seibel E M, Gibson Q D, Xie W and Cava R J 2015 *APL Materials* **3** 083602

[76] Bian G, Chang T R, Zheng H, Velury S, Xu S Y, Neupert T, Chiu C K, Huang S M, Sanchez D S, Belopolski I, Alidoust N, Chen P J, Chang G, Bansil A, Jeng H T, Lin H and Hasan M Z 2016 *Phys. Rev. B* **93** 121113

[77] Hu J, Tang Z, Liu J, Liu X, Zhu Y, Graf D, Myhro K, Tran S, Lau C N, Wei J and Mao Z 2016 *Phys. Rev. Lett.* **117** 016602

[78] Neupane M, Belopolski I, Hosen M M, Sanchez D S, Sankar R, Szlawska M, Xu S, Dimitri K, Dhakal N, Maldonado P, Oppeneer P M, Kaczorowski D, Chou F, Hasan M Z and Durakiewicz T 2016 *Phys. Rev. B* **93** 201104

[79] Schoop L M, Ali M N, Straßer C, Topp A, Varykhalov A, Marchenko D, Duppel V, Parkin S S, Lotsch B V and Ast C R 2016 *Nat. Commun.* **7** 11696

[80] Wu Y, Wang L L, Mun E, Johnson D D, Mou D X, Huang L N, Lee Y J, Bud'ko S L, Canfield P C and Kaminski A 2016 *Nat. Phys.* **12** 667

[81] Feng B, Fu B, Kasamatsu S et al. 2017 *Nat. Commun.* **8** 1007

[82] Hosur P and Qi X 2013 *C. R. Phys.* **14** 857

[83] He L P and Li S Y 2016 *Chin. Phys. B* **25** 117105

[84] Rao S arXiv:1603.02821

[85] Lu H Z and Shen S Q 2016 *Chin. Phys. B* **25** 117202

[86] Jia S, Xu S Y and Hasan M Z 2016 *Nat. Mater.* **15** 5 1140

[87] Lu H Z and Shen S Q 2017 *Front. Phys.* **12**3 127201

[88] Song H D, Sheng D, Wang A Q, Li J G, Yu D P and Liao Z M 2017 *Chin. Phys. B* **26** 037301

[89] Dai X 2016 *Nat. Mater.* **15** 5

[90] Ali M N, Gibson Q, Jeon S, Zhou B B, Yazdani A and Cava R J. 2014 *Inorg Chem* **53** 4062

[91] Abrikosov A A 1998 *Phys. Rev. B* **58** 2788

[92] Narayanan A, Watson M D, Blake S F, Bruyant N, Drigo L, Chen Y L, Prabhakaran D, Yan B, Felser C, Kong T, Canfield P C and Coldea A I 2015 *Phys. Rev. Lett.* **114** 117201

[93] Murakawa H, Bahramy M S, Tokunaga M, Kohama Y, Bell C, Kaneko Y, Nagaosa N, Hwang H Y and Tokura Y 2013 *Science* **342** 1490

[94] Wang C M, Lu H Z and Shen S Q 2016 *Phys. Rev. Lett.* **117** 077201

[95] Liu Y W , Zhang C , Yuan X , Lei T , Wang C , Di Sante D , Narayan A , He L , Picozzi S , Sanvito S , Che R C and Xiu F X 2015 *NPG Asia Mater.* **7** e221

[96] Zhao B, Cheng P, Pan H, Zhang S, Wang B, Wang G, Xiu F and Song F 2016 *Sci. Rep.* **6** 22377

[97] Li C Z, Zhu R, Ke X, Zhang J M, Wang L X, Zhang L, Liao Z M and Yu D P 2015 *Cryst Growth Des.* **15** 3264

[98] Chen Z G, Zhang C, Zou Y, Zhang E, Yang L, Hong M, Xiu F and Zou J 2015 *Nano Lett.* **15** 5830

[99] Zhang E, Liu Y, Wang W, Zhang C, Zhou P, Chen Z G, Zou J and Xiu F 2015 *ACS Nano* **9** 8843

[100] Li C Z, Wang L X, Liu H, Wang J, Liao Z M and Yu D P 2015 *Nat. Commun.* **6** 10137

[101] Li H, He H, Lu H Z, Zhang H, Liu H, Ma R, Fan Z, Shen S Q and Wang J 2016 *Nat. Commun.* **7** 10301

[102] Zhang C, Zhang E, Wang W, Liu Y, Chen ZG, Lu S, Liang S, Cao J, Yuan X, Tang L, Li Q, Zhou C, Gu T, Wu Y, Zou J and Xiu F 2017 *Nat. Commun.* **8**

13741

[103] Li C Z, Li J G, Wang L X, Zhang L, Zhang J M, Yu D and Liao Z M 2016 *ACS Nano* **10** 6020

[104] Cheng P, Zhang C, Liu Y, Yuan X, Song F, Sun Q, Zhou P, Zhang D W and Xiu F 2016 *New J. Phys.* **18** 083003

[105] Yuan X, Cheng P, Zhang L, Zhang C, Wang J, Liu Y, Sun Q, Zhou P, Zhang DW, Hu Z, Wan X, Yan H, Li Z and Xiu F 2017 *Nano Lett.* **17** 2211

[106] Wang L X, Li C Z, Yu D P and Liao Z M 2016 *Nat. Commun.* **7** 10769

[107] Wang L X, Wang S, Li J G, Li C Z, Yu D P and Liao Z M 2016 *Phys. Rev. B* **94** 161402(R)

[108] Wang L X, Wang S, Li J G, Li C Z, Xu J, Yu D P and Liao Z M 2017 *J. Phys. Condens. Matter* **29** 044003

[109] Cao J, Liang S, Zhang C, Liu Y, Huang J, Jin Z, Chen Z G, Wang Z, Wang Q, Zhao J, Li S, Dai X, Zou J, Xia Z, Li L and Xiu F 2015 *Nat. Commun.* **6** 7779

[110] Adler S L 1969 *Phys. Rev.* **177** 2426

[111] Bell J S and Jackiw R A 1969 *Il Nuovo Cimento A* **60** 47

[112] Nielsen H B and Ninomiya M 1983 *Phys. Lett. B* **130** 389

[113] Burkov A A 2015 *J. Phys. Condens. Matter* **27** 113201

[114] Burkov A A 2015 *Phys. Rev. B* **91** 245157

[115] Son D T and Spivak B Z 2013 *Phys. Rev. B* **88** 104412

[116] Andrew C, Potter I K, Ashvin Vishwanath 2014 *Nat. Commun.* **5**

[117] Moll PJ, Nair N L, Helm T, Potter A C, Kimchi I, Vishwanath A and Analytis J G 2016 *Nature* **535** 266

[118] Kargarian M, Randeria M and Lu Y M 2016 *Proc. Natl. Acad. Sci.* **113** 8648

[119] Zheng G, Wu M, Zhang H, Chu W, Gao W, Lu J, Han Y, Yang J, Du H, Ning W, Zhang Y and Tian M 2017 *Phys. Rev. B* **96** 121407(R)

[120] Zhang C, Narayan A, Lu S, Zhang J, Zhang H, Ni Z, Yuan X, Liu Y, Park J H, Zhang E, Wang W, Liu S, Cheng L, Pi L, Sheng Z, Sanvito S and Xiu F 2017 *Nat. Commun.* **8** 1272

[121] Schumann T, Galletti L, Kealhofer DA, Kim H, Goyal M and Stemmer S 2018 *Phys. Rev. Lett.* **120** 016801

[122] Uchida M, Nakazawa Y, Nishihaya S, Akiba K, Kriener M, Kozuka Y, Miyake A, Taguchi Y, Tokunaga M, Nagaosa N, Tokura Y and Kawasaki M 2017 *Nat. Commun.* **8** 2274

[123] Alicea J 2012 *Rep. Prog. Phys.* **75** 076501

[124] Ryu S, Schnyder A P, Furusaki A and Ludwig A W W 2010 *New J. Phys.* **12** 065010

[125] Fu L and Kane C L 2008 *Phys. Rev. Lett.* **100** 096407

[126] Read N and Green D 2000 *Phys. Rev. B* **61** 10267

[127] Kitaev A Y 2001 *Physics Uspekhi* **44** 131

[128] Ivanov D A 2001 *Phys. Rev. Lett.* **86** 268

[129] Nayak C, Simon S H, Stern A, Freedman M and Sarma S D 2008 *Rev. Mod. Phys.* **80** 1083

[130] Alicea J, Oreg Y, Rafael G, Oppen F and Fisher M P A 2011 *Nat. Phys.* **7** 412

[131] Wang H, Wang H, Liu H, Lu H, Yang W, Jia S, Liu X J, Xie X C, Wei J and Wang J 2016 *Nat. Mater.* **15** 38

[132] Kobayashi S and Sato M 2015 *Phys. Rev. Lett.* **115** 187001

[133] Daghero D and Gonnelli R S 2010 *Supercond. Sci. Technol.* **23** 043001

[134] Deutscher G 2005 *Rev. Mod. Phys.* **77** 109

[135] He L, Jia Y, Zhang S, Hong X, Jin C and Li S 2016 n*pj Quantum Materials* **1** 16014

[136] Aggarwal L, Gaurav A, Thakur G S, Haque Z, Ganguli A K and Sheet G 2016 *Nat. Mater.* **15** 32

[137] Xiong J, Kushwaha S, Krizan J, Liang T, Cava R J and Ong N P 2016 *Europhys. Lett.* **114** 27002

[138] Hellerstedt J, Yudhistira I, Edmonds M T, Liu C, Collins J, Adam S and Fuhrer M S 2017 *Phys. Rev. Materials* **1** 054203

[139] Gao W, Hao N, Zheng F W, Ning W, Wu M, Zhu X, Zheng G, Zhang J, Lu J, Zhang H, Xi C, Yang J, Du H, Zhang P, Zhang Y and Tian M 2017 *Phys. Rev. Lett.* **118** 256601

[140] Weng H, Dai X and Fang Z 2014 *Phys. Rev. X* **4**

[141] Li Q, Kharzeev D E, Zhang C, Huang Y, Pletikosic I, Fedorov A V, Zhong R D, Schneeloch J A, Gu G D and Valla T 2016 *Nat. Phys.* **12** 550

[142] Chen R Y, Zhang S J, Schneeloch J A, Zhang C, Li Q, Gu G D and Wang N L 2015 *Phys. Rev. B* **92** 075107

[143] Chen R Y, Chen Z G, Song X-Y, Schneeloch J A, Gu G D, Wang F and Wang N L 2015 *Phys. Rev. Lett.* **115** 176404

[144] Zheng G, Lu J, Zhu X, Ning W, Han Y, Zhang H, Zhang J, Xi C, Yang J, Du H,

Yang K, Zhang Y and Tian M 2016 *Phys. Rev. B* **93** 115414

[145] Yuan X, Zhang C, Liu Y, Narayan A, Song C, Shen S, Sui X, Xu J, Yu H, An Z, Zhao J, Sanvito S, Yan H and Xiu F 2016 *NPG Asia Mater.* **8** e325

[146] Liu Y, Yuan X, Zhang C, Jin Z, Narayan A, Luo C, Chen Z, Yang L, Zou J, Wu X, Sanvito S, Xia Z, Li L, Wang Z and Xiu F 2016 *Nat. Commun.* **7** 12516

[147] Wang H C, Liu H W, Li Y N, Liu Y J, Wang J F, Liu J, Wang Y, Li L, Yan J Q, Mandrus D, Xie X C and Wang J arXiv:1704.00995

[148] Yu W, Jiang Y, Yang J, Dun Z L, Zhou H D, Jiang Z, Lu P and Pan W 2016 *Sci. Rep.* **6** 35357

[149] Zhou Y, Wu J, Ning W, Li N, Du Y, Chen X, Zhang R, Chi Z, Wang X, Zhu X, Lu P, Ji C, Wan X, Yang Z, Sun J, Yang W, Tian M, Zhang Y and Mao H K 2016 *Proc. Natl. Acad. Sci.* **113** 2904

[150] Zhang J L, Guo C Y, Zhu X D, Ma L, Zheng G L, Wang Y Q, Pi L, Chen Y, Yuan H Q and Tian M L 2017 *Phys. Rev. Lett.* **118** 206601

[151] Zheng G, Zhu X, Liu Y, Lu J, Ning W, Zhang H, Gao W, Han Y, Yang J, Du H, Yang K, Zhang Y and Tian M 2017 *Phys. Rev. B* **96** 121401(R)

[152] Lu J, Zheng G, Zhu X, Ning W, Zhang H, Yang J, Du H, Yang K, Lu H, Zhang Y and Tian M 2017 *Phys. Rev. B* **95** 125135

[153] Niu J, Wang J, He Z, Zhang C, Li X, Cai T, Ma X, Jia S, Yu D and Wu X 2017 *Phys. Rev. B* **95** 035420

[154] Shahi P, Singh D J, Sun J P, Zhao L X, Chen G F, Yan J Q, Mandrus D G and Cheng J G arXiv:1611.06370

[155] Qiu G, Du Y, Charnas A, Zhou H, Jin S, Luo Z, Zemlyanov D Y, Xu X, Cheng G J and Ye P D 2016 *Nano Lett.* **16** 7364

[156] Hang C 2017 *New J. Phys.* **19** 015005

[157] Pariari A and Mandal P 2017 *Sci. Rep.* **7** 40327

[158] Liang T, Lin J, Gibson Q, Kushwaha S, Liu M, Wang W, Xiong H, Sobota J A, Hashimoto M, Kirchmann P S, Shen Z X, Cava R J and Ong N P 2018 *Nat. Phys.* **14**, 451

[159] Lv Y Y, Zhang F, Zhang B B, Pang B, Yao S H, Chen Y B, Ye L W, Zhou J, Zhang S T, Chen Y F 2017 *Journal of Crystal Growth* **457** 250

[160] Li X B, Huang W K, Lv YY, Zhang K W, Yang C L, Zhang B B, Chen Y B, Yao S H, Zhou J, Lu M H, Sheng L, Li S C, Jia J F, Xue Q K, Chen Y F and Xing D Y 2016 *Phys. Rev. Lett.* **116** 176803

[161] Wu R, Ma J Z, Nie S M, Zhao L X, Huang X, Yin J X, Fu B B, Richard P, Chen G F, Fang Z, Dai X, Weng H M, Qian T, Ding H and Pan S H 2016 *Phys. Rev. X* **6** 021017

[162] Zhang Y, Wang C, Yu L et al. 2017 *Nat. Commun.* **8** 15512

[163] Manzoni G, Gragnaniello L, Autès G, Kuhn T, Sterzi A, Cilento F, Zacchigna M, Enenkel V, Vobornik I, Barba L, Bisti F, Bugnon P, Magrez A, Strocov VN, Berger H, Yazyev OV, Fonin M, Parmigiani F and Crepaldi A 2016 *Phys. Rev. Lett.* **117** 237601

[164] Jiang Y, Dun Z L, Zhou H D, Lu Z, Chen K W, Moon S, Besara T, Siegrist T M, Baumbach R E, Smirnov D and Jiang Z 2017 *Phys. Rev. B* **96** 041101

[165] Chen Z G, Chen R Y, Zhong R D, Schneeloch J, Zhang C, Huang Y, Qu F, Yu R, Li Q, Gu G D and Wang N L 2017 *Proc. Natl. Acad. Sci.* **114** 816

[166] Fan Z, Liang Q F, Chen Y B, Yao S H and Zhou J 2017 *Sci. Rep.* **7** 45667

[167] McIlroy D N, Moore S, Zhang D, Wharton J, Kempton B, Littleton R, Wilson M, Tritt T M and Olson C G 2004 *J. Phys. Condens. Matter* **16** L359

[168] Tritt T M, Lowhorn N D, Littleton IV R T, Pope A, Feger C R and Kolis J W 1999 *Phys. Rev. B* **60** 7816

[169] Okada S, Sambongi T and Ido M 1980 *J. Phys. Soc. Jpn.* **49** 839

[170] Izumi M, Uchinokura K and Matsuura E 1981 *Solid State Commun.* **37** 641

[171] Shaviv R, Westrum E F, Fjellvåg H and Kjekshus A 1989 *J. Solid State Chem.* **81** 103

[172] Kamm G, Gillespie D, Ehrlich A, Peebles D and Levy F 1987 *Phys. Rev. B* **35** 1223

[173] Izumi M, Nakayama T, Uchinokura K, Harada S, Yoshizaki R and Matsuura E 1987 *J. Phys. C* **20** 3691

[174] Wang H C, Li C K, Liu H W, Yan J Q, Wang J F, Liu J, Lin Z Q, Li Y N, Wang Y, Li L, Mandrus D, Xie X C, Feng J and Wang J 2016 *Phys. Rev. B* **93** 165127

[175] Zhao L X, Huang X C, Long Y J, Chen D, Liang H, Yang Z H, Xue M Q, Ren Z A, Weng H M, Fang Z, Dai X, Chen G F 2017 *Chin. Phys. Lett.* **34** 037102

[176] Qi Y, Shi W, Naumov P G, Kumar N, Schnelle W, Barkalov O, Shekhar C, Borrmann H, Felser C, Yan B and Medvedev S A 2016 *Phys. Rev. B* **94** 054517

[177] Liu Y, J Long Y, Zhao L X, Nie S M, Zhang S J, Weng Y X, JinM L, Li W M, Liu Q Q, Long Y W, Yu R C, Fen X L, Li Q, Weng H M, Dai X, Fang Z, Chen G F and Jin C Q arXiv:1603.00514

[178] Yan M, Huang H, Zhang K, Wang E, Yao W, Deng K, Wan G, Zhang H, Arita M, Yang H, Sun Z, Yao H, Wu Y, Fan S, Duan W and Zhou S 2017 *Nat. Commun.* **8** 257

[179] Andreev A V and Spivak B Z 2018 *Phys. Rev. Lett.* **120** 026601

[180] Dai X, Du Z Z and Lu H Z 2017 *Phys. Rev. Lett.* **119** 166601

[181] Wang J, Li H, Chang C, He K, Lee J S, Lu H, Sun Y, Ma X, Samarth N, Shen S, Xue Q K, Xie M H and Chan M H W 2012 *Nano Res.* **5** 739

[182] Zhang C L, Tong B B, Yuan Z J, Lin Z Q, Wang J F, Zhang J L, Xi C Y, Wang Z, Jia S and Zhang C 2016 *Phys. Rev. B* **94** 205120

[183] Zhang Q R, Zeng B, Rhodes D, Memaran S, Besara T, Sankar R, Chou F, Alidoust N, Xu S Y, Belopolski I, Hasan M Z and Balicas L arXiv:1705.00920

[184] Ramshaw B J, Modic K A, Shekhter A, Zhang Y, Kim E A, Moll P J W, Bachmann M, Chan M K, Betts J B, Balakirev F, Migliori A, Ghimire N J, Bauer E D, Ronning F and McDonald R D 2018 *Nat. Commun.* **9**, 2217

[185] Moll P J, Potter A C, Nair N L, Ramshaw B J, Modic K A, Riggs S, Zeng B, Ghimire N J, Bauer E D, Kealhofer R, Ronning F and Analytis J G 2016 *Nat. Commun.* **7** 12492

[186] Zhang C L, Xu S Y, Wang C M et al. 2017 *Nat. Phys.* **13** 979

[187] Brignall N L 1974 *J Phys C* **7** 4266

[188] Song Z, Fang Z and Dai X 2017 *Phys. Rev. B* **96** 235104

[189] Wang H, Wang H C, Chen Y Q, Luo J W, Yuan Z J, Liu J, Wang Y, Jia S, Liu X J, Wei J and Wang J 2017 *Chin Sci Bull* **62** 425

[190] Zhou Y, Lu P, Du Y, Zhu X, Zhang G, Zhang R, Shao D, Chen X, Wang X, Tian M, Sun J, Wan X, Yang Z, Yang W, Zhang Y and Xing D 2016 *Phys. Rev. Lett.* **117** 146402

[191] Soluyanov A A, Gresch D, Wang Z, Wu Q, Troyer M, Dai X and Bernevig B A 2015 *Nature* **527** 495

[192] Chang T R, Xu S Y, Chang G, Lee C C, Huang S M, Wang B, Bian G, Zheng H, Sanchez D S, Belopolski I, Alidoust N, Neupane M, Bansil A, Jeng H T, Lin H and Hasan M Z 2016 *Nat. Commun.* **7** 10639

[193] Sun Y, Wu S C, Ali M N, Felser C and Yan B 2015 *Phys. Rev. B* **92** 161107(R)

[194] Wang Z, Gresch D, Soluyanov A A, Xie W, Kushwaha S, Dai X, Troyer M, Cava R J and Bernevig B A 2016 *Phys. Rev. Lett.* **117** 056805

[195] Deng K, Wan G, Deng P, Zhang K, Ding S, Wang E, Yan M, Huang H, Zhang

H, Xu Z, Denlinger J, Fedorov A, Yang H, Duan W, Yao H, Wu Y, Fan S, Zhang H, Chen X and Zhou S 2016 *Nat. Phys.* **12** 1105

[196] Huang L, McCormick T M, Ochi M, Zhao Z, Suzuki M T, Arita R, Wu Y, Mou D, Cao H, Yan J, Trivedi N and Kaminski A 2016 *Nat. Mater.* **15** 1155

[197] Tamai A, Wu Q S, Cucchi I, Bruno F Y, Riccò S, Kim T K, Hoesch M, Barreteau C, Giannini E, Besnard C, Soluyanov A A and Baumberger F 2016 *Phys. Rev. X* **6** 031021

[198] Liang A, Huang J, Nie S et al. arXiv:1604.01706

[199] Jiang J, Liu Z K, Sun Y, Yang H F, Rajamathi C R, Qi Y P, Yang L X, Chen C, Peng H, Hwang C C, Sun S Z, Mo S K, Vobornik I, Fujii J, Parkin S S, Felser C, Yan B H and Chen Y L 2017 *Nat. Commun.* **8** 13973

[200] Qi Y, Naumov P G, Ali M N et al. 2016 *Nat. Commun.* **7** 11038

[201] Chen F C, Luo X, Xiao R C, Lu W J, Zhang B, Yang H X, Li J Q, Pei Q L, Shao D F, Zhang R R, Ling L S, Xi C Y, Song W H and Sun Y P 2016 *Appl. Phys. Lett.* **108** 162601

[202] Li Y N, Gu Q Q, Chen C., Zhang J, Liu Q, Hu X Y, Liu J, Liu Y, Ling L S, Tian M L, Wang Y, Samarth N, Li S Y, Zhang T, Feng J and Wang J *DOI: 10.1073/pnas.1801650115*

[203] Guguchia Z, Rohr F, Shermadini Z et al. 2017 *Nat Commun* **8** 1082

[204] Ali M N, Xiong J, Flynn S, Tao J, Gibson Q D, Schoop L M, Liang T, Haldolaarachchige N, Hirschberger M, Ong N P and Cava R J 2014 *Nature* **514** 205

[205] Pletikosić I, Ali M N, Fedorov A V, Cava R J and Valla T 2014 *Phys. Rev. Lett.* **113** 216601

[206] Qian X, Liu J, Fu L and Li J 2014 *Science* **346** 1344

[207] Cai P L, Hu J, He L P, Pan J, Hong X C, Zhang Z, Zhang J, Wei J, Mao Z Q and Li S Y 2015 *Phys. Rev. Lett.* **115** 057202

[208] Kang D, Zhou Y, Yi W, Yang C, Guo J, Shi Y, Zhang S, Wang Z, Zhang C, Jiang S, Li A, Yang K, Wu Q, Zhang G, Sun L and Zhao Z 2015 *Nat. Commun.* **6** 7804

[209] Lv H Y, Lu W J, Shao D F, Liu Y, Tan S G and Sun Y P 2015 *Europhys Lett.* **110** 37004

[210] Pan X C, Chen X, Liu H, Feng Y, Wei Z, Zhou Y, Chi Z, Pi L, Yen F, Song F, Wan X, Yang Z, Wang B, Wang G, and Zhang Y 2015 *Nat. Commun.* **6** 7805

[211] Zhao Y F, Liu H W, Yan J Q, An W, Liu J, Zhang X, Wang H C, Liu Y, Jiang H, Li Q, Wang Y, Li X Z, Mandrus D, Xie X C, Pan M H and Wang J 2015 *Phys. Rev. B* **92** 041104

[212] Wang C, Zhang Y, Huang J et al. 2016 *Phys. Rev. B* **94** 241119

[213] Wang Y, Liu E, Liu H, Pan Y, Zhang L, Zeng J, Fu Y, Wang M, Xu K, Huang Z, Wang Z, Lu H Z, Xing D, Wang B, Wan X and Miao F 2016 *Nat. Commun.* **7** 13142

[214] Lv Y Y, Li X, Zhang B B, Deng W Y, Yao S H, Chen Y B, Zhou J, Zhang S T, Lu M H, Zhang L, Tian M, Sheng L and Chen Y F 2017 *Phys. Rev. Lett.* **118** 096603

[215] Fang C, Gilbert M J, Dai X and Bernevig B A 2012 *Phys. Rev. Lett.* **108** 266802

[216] Xu G, Weng H, Wang Z, Dai X and Fang Z. 2011 *Phys. Rev. Lett.* **107** 186806

[217] Huang S M, Xu S Y, Belopolski I, Lee C C, Chang G, Chang T R, Wang B, Alidoust N, Bian G, Neupane M, Sanchez D, Zheng H, Jeng H T, Bansil A, Neupert T, Lin H and Hasan M Z 2016 *Proc. Natl. Acad. Sci.* **113** 1180

[218] Jian S K and Yao H 2015 *Phys. Rev. B* **92** 045121

[219] Lai H H 2015 *Phys. Rev. B* **91** 235131

[220] Burkov A A 2016 *Nat. Mater.* **15** 1145

[221] Hirschberger M 2016 *Nat. Mater.* **15** 1161

[222] Felser C and Yan B 2016 *Nat. Mater.* **15** 1149

[223] Liu J Y, Hu J, Zhang Q, Graf D, Cao H B, Radmanesh S M A, Adams D J, Zhu Y L, Cheng G F, Liu X, Phelan W A, Wei J, Jaime M, Balakirev F, Tennant D A, DiTusa J F, Chiorescu I, Spinu L and Mao Z Q 2017 *Nat. Mater.* **16** 905

[224] Wang Y J, Yu J H, Wang Y Q, Xi C Y, Ling L S, Zhang S L, Wang J R, Xiong Y M, Han T, Han H, Yang J, Gong J X, Luo L, Tong W, Zhang L, Qu Z, Han Y Y, Zhu W K, Pi L, Wan X G, Zhang C J and Zhang Y H 2018 *Phys Rev B* **97** 115133

[225] Bradlyn B, Cano J, Wang Z, Vergniory M G, Felser C, Cava R J and Bernevig B A 2016 *Science* **353** 6299

[226] Weng H, Fang C, Fang Z and Dai X 2016 *Phys. Rev. B* **94** 165201

[227] Weng H, Fang C, Fang Z and Dai X 2016 *Phys. Rev. B* **93** 241202

[228] Chang G, Xu S Y, Huang S M, Sanchez D S, Hsu C H, Bian G, Yu Z M, Belopolski I, Alidoust N, Zheng H, Chang T R, Jeng H T, Yang S A, Neupert T, Lin H and Hasan M Z 2017 *Sci. Rep.* **7** 1688

[229] Lv B Q, Feng Z L, Xu Q N, Gao X, Ma J Z, Kong L Y, Richard P, Huang Y B, Strocov V N, Fang C, Weng H M, Shi Y G, Qian T and Ding H 2017 *Nature* **546** 627

[230] Ma J Z, He J B, Xu Y F et al. 2018 *Nat. Phys.* **14**, 349

[231] He J B, Chen D, Zhu W L, Zhang S, Zhao L X, Ren Z A and Chen G F 2017 *Phys. Rev. B* **95** 195165

[232] Shekhar C, Sun Y, Kumar N, Nicklas M, Manna K, Suess V, Young O, Leermakers I, Foerster T, Schmidt M, Muechler L, Werner P, Schnelle W, Zeitler U, Yan B, Parkin S S P and Felser C arXiv:1703.03736

[233] Lu W, Ge S F, Liu X F, Lu H, Li C Z, Lai J W, Zhao C, Liao Z M, Jia S and Sun D 2016 *Phys. Rev. B* **95** 024303

[234] Wang Q, Li C Z, Ge S, Li J G, Lu W, Lai J, Liu X, Ma J, Yu D P, Liao Z M and Sun D 2017 *Nano Lett.* **17** 834

[235] Zhu C, Wang F, Meng Y, Yuan X, Xiu F, Luo H, Wang Y, Li J, Lv X, He L, Xu Y, Liu J, Zhang C, Shi Y, Zhang R and Zhu S 2017 *Nat. Commun.* **8** 14111

[236] Jia Z, Li C, Li X, Shi J, Liao Z, Yu D and Wu X 2016 *Nat. Commun.* **7** 13013

[237] Zhou T, Zhang C, Zhang H, Xiu F and Yang Z 2016 *Inorg. Chem. Frontiers* **3** 1637